\documentclass[graybox,secnum]{svmult}

\usepackage{mathptmx}       
\usepackage{helvet}         
\usepackage{courier}        
\usepackage{type1cm}        
\usepackage{makeidx}         
\usepackage{graphicx}        
\usepackage{multicol}        
\usepackage[bottom]{footmisc}

\usepackage{hyperref}        
\usepackage{soul}            
\hypersetup{colorlinks=true,urlcolor=blue}
\usepackage[square,numbers]{natbib}
 
\usepackage{newtxtext} 
\usepackage{newtxmath}       
\usepackage{wasysym}

\makeindex             

\begin{document}

\title*{Star Forming Regions}
\author{S. Sciortino}
\institute{Salvatore Sciortino \at INAF-Osservatorio Astronomico di Palermo, Piazza del Parlamento 1, 90138 Palermo, Sicily, Italy, \email{salvatore.sciortino@inaf.it}
}
%
%
\maketitle

\abstract*{Since the '80s the \textit{Einstein} observatory has shown the Young Stellar Objects (YSOs), emit X-rays with luminosities, in the 0.3--8 keV bandpass, up to $\rm 10^3$--$\rm 10^4$ times than the Sun and that the X-ray emission is highly variable. ROSAT has confirmed the pervasiveness of X-ray emission from YSOs and ASCA has provided evidence that the emission of Class I YSOs is largely originating from optical thin-plasma at temperature of 1-50 $\times 10^6$ K. These intrinsic, unexpected, properties and the transformational capabilities of the \textit{Chandra} and \textit{XMM-Newton} observatories has made X-rays a powerful tool to trace the star formation process up to distance of a few kpc around the Sun. Starting from the early evidences of the '80s and the intriguing questions they raised, I will summarize the results obtained and how they have influenced our current understanding of physical processes at work and I will discuss some of the still open issues and some of the likely avenues that next generation X-ray observatory will open.}

\abstract{Since the '80s the \textit{Einstein} observatory has shown that the Young Stellar Objects (YSOs), emit X-rays with luminosities, in the 0.3--8 keV bandpass, up to $\rm 10^3$--$\rm 10^4$ times than the Sun and that the X-ray emission is highly variable. ROSAT has confirmed the pervasiveness of X-ray emission from YSOs and ASCA has provided evidence that the emission of Class I YSOs is largely originating from optical thin-plasma at temperature of 1-50 $\times 10^6$ K. These intrinsic, unexpected, properties and the transformational capabilities of the \textit{Chandra} and \textit{XMM-Newton} observatories has made X-rays a powerful tool to trace the star formation process up to distance of a few kpc around the Sun. Starting from the early evidences of the '80s and the intriguing questions they raised, I will summarize the results obtained and how they have influenced our current understanding of physical processes at work and I will discuss some of the still open issues and some of the likely avenues that next generation X-ray observatory will open.}


\section{Introduction}
\label{sec:1}
The advent of imaging X-ray astronomy with telescopes covering not too-narrow field of view (FOV), photon counting detectors providing time, spectral and angular resolutions has had and will continue to have a great impact on the studies of star forming regions (SFRs). Since the '80s the spectral and angular resolutions have increased going from $\rm \Delta E/E \sim$ 1 at 1.5 keV (with the \textit{Einstein} Imaging Proportional Counter \cite{Giacconi+1979Einstein} ) to $\sim$ 0.1 at 1.5 keV (\textit{Chandra} ACIS or \textit{XMM-Newton} EPIC) and from $\sim$ 1 arc-min to $\sim$ 10 arc-sec (\textit{XMM-Newton}) or even to $\sim$ 0.5 arc-sec (\textit{Chandra} FOV center). This increase of observational capabilities 
has allowed us to start answering questions on the physical processes regulating the interaction of the still contracting star and its surrounding disk, on the effect of the "local" environment on the evolution of circumstellar disks, as well as on the processes regulating the star formation (recently reviewed by \cite{Feigelson2018}).     

The key role of X-ray observations in the SFR studies can be appreciated by considering that: i) the YSOs do emit copious amount of X-rays; ii) the optical depth of the $\sim$ 1 keV X-rays is comparable or smaller than the optical depth in the IR 
(cf. \cite{Casanova+1995rhoOph}), 
hence X-ray observations can really peer deep inside the star forming sites, finding the vast population of X-ray emitting YSOs and allowing to derive a census of their stellar populations as unbiased as possible; iii) the complex physical processes at work imprint characteristic signatures on the X-ray spectra and light-curves. These can be recognized and investigated as a part of a multi-wavelength studies;
iv) the YSOs are clearly clustered in specific region of the sky, the physical extent of each individual cluster/group typically is 1--10 pc corresponding to apparent angular size ranging from 0.5--5 sq. degree for the nearest SFRs at $\sim$ 130 pc, to less than 0.5 sq. degree already at 1 kpc. Such an extent matches the FOV of the past and current imaging X-ray observatories\footnote{The case of the nearby and dispersed SFRs, like Taurus-Auriga, covering a wider area of the sky, can be and has been addressed either with multiple observations or with the ROSAT all-sky survey}; v) several detailed predictions indicate the YSO X-rays as a major "local" ionization source with likely a key role in regulating the overall star formation process \cite{LorenzaniPalla2001, Locci+2018MNRAS}, as well as a key ingredient in the early evolution of the proto-planetary disk (e.g.\cite{Adams+2012PASP}).

In the case of intermediate and low-mass YSO four major phases have been recognized, essentially by the slope of the IR to millimetre spectral energy distribution 
\cite{AndreMontmerle1994} : Class~0 sources (still accreting the bulk of their material, emit most at mm wavelengths, $age \rm \sim 10^4$ yr);  Class~I (proto-star with strong circumstellar disk, spectra dominated by far-IR emission from in-falling envelope, $age \rm \sim 10^5$ yr); Class~II (Classical T-Tauri phase, $age \rm \sim 10^6$--$\rm 10^7$ yr) and Class~III (Weak lined T Tauri star, $age \rm \sim 10^6$--$\rm 10^7$ yr). The so-called "flat spectrum" sources lie at the Class~I/Class~II boundary. A schematic summary, including the presence of thermal or non-thermal radio emission, is provided in Fig. 1 of \cite{FeigelsonMontmerle1999ARAA}. X-ray emission from the Class~0 YSOs has been searched over many decades but the efforts have been hampered being deeply embedded in their parental molecular clouds, surrounded by in-falling envelopes as well as by out-flowing material (jets), and may be even self-shadowed by their own circumstellar disks. A$_V$ ranges from 10 to 50 in Class~I YSOs and may reach $\sim$ 1000 for Class~0 YSOs. While firm evidence of X-ray emission in Class~I YSO has been found since the '90s (see Sect. \ref{sec:4}), only recently evidence of X-ray emission from Class~0 YSOs has been reported (see Sect. \ref{sec:5.3}). 
In the case of massive stars an established analogous observational evolutionary sequence lacks, even if we know  
the crucial role played by their radiation field since formation phase (\cite{Motte+2018ARAA} and references therein).
 
In massive stars X-rays are thought to be generated from the natural instabilities of the line-driven stellar wind \cite{LucyWhite1980ApJ, CassinelliSwank1983ApJ, Feldmeier+1997AandA} even if, in some cases, evidence for, a likely additional, component of magnetic origin has been found (e.g. \cite{Gagne+2011OB}). In massive close binary systems an additional component due to wind-wind interaction is clearly present and required to explain the X-ray variability at various orbital phases (for a recent review see \cite{RauwNaze2016review} and other contributions in this Handbook section). Available data are consistent with a rapid settling of the X-ray emission from massive stars\footnote{The X-ray imaging surveys of massive SFRs in the Galaxy have shown to be quite effective also in improving the census of the massive stars that is still incomplete because of high extinction, bright diffuse optical and IR emission in H II regions, and field star contamination (e.g. \cite{Povich+2017ApJ}).}.

The origin of the X-ray emission of Herbig Ae/Be stars, namely the intermediate mass nearly fully radiative counterparts of the classical T Tauri stars, is still enigmatic. Their internal structure should not support a dynamo action like the one of lower mass YSOs hence it is disputed if X-rays traces a magnetic activity or if they are produced by other mechanisms. 
When those stars arrive on the main sequence they have $L_X$ \rm $\sim$ 10$^{25}$ erg/sec 
hence the about six order of magnitude decline of $L_X$  associated to the Hergig Ae/Be phase and related to the presence of disks, accretion, or associated jet-type activity has to be very rapid.  Radiative winds are ruled out as the main emission mechanism on the basis of the high X-ray temperatures of emitting plasma, while results on the low-mass companion hypothesis are not firmly conclusive \cite{Stelzer+2006HAeBe, Stelzer+2009HAeBe}.

Since X-ray emission from massive stars settles on rapidly, one important, and controversial, issue, is the effect that the high-energy emission of the massive SFR members has on the evolution of circumstellar disks of the low mass YSO population (see Sect. \ref{sec:5.5}). Refined numerical models have shown the essential role of the magnetic field on the birth of 
the proto-planetary disk (e.g. \cite{Lebreuilly+2021ApJ}).

Given space limitation and review focus I will only briefly mention the diffuse X-ray emission (e.g. \cite{Townsley+2011a, Townsley+2011b, Townsley+2019}) found in all massive SFRs as well as the many results from high resolution X-ray spectroscopy. This latter subject has been reviewed by \cite{GuedelNaze2009AandARv, Argiroffi2019AN} as well as in other contributions in this Handbook section. 

Since the X-ray investigations of star forming regions have impressively grown in the last couple of decades, I can account only for a very
small subset of them. Selection has been done to illustrate specific astrophysical questions through few representative cases and/or major achievements. I have adopted as a "fil rouge" the improvement of observational capabilities over the last 40 years and the new frontiers they have opened. While the X-ray observations have a pivotal role only a multi-wavelength approach, including optical, IR and mm data, allow us to address the complex physics at work.
 
YSO X-ray studies before \textit{Chandra} and \textit{XMM-Newton} has been extensively reviewed elsewhere \cite{FeigelsonMontmerle1999ARAA}.  More recent results have been reviewed over the time in various occasions \cite{Feigelson2010PNAS, Feigelson2018, Montmerle2007IAUS, Gunther2013AN, Argiroffi2019AN, Sciortino+2019AN}.
The many facets of the star formation have been presented in an organized form in a reference book \cite{StahlerPalla2004book} 
as well as in two monographs, one more focused on the accretion process\citep{Hartmann2008book}, and the other with a specific emphasis on the role of high energy processes \citep{Schulz2012book}.

\section{The early Einstein discoveries, the emergence of intriguing questions and some initial answers}
\label{sec:2}
The first, indeed very sparse, evidence of YSO X-ray emission in a SFR dates back to 1978.  
The emission from the source 3U 0527-05 was found with the Astronomical Netherlands Satellite to come from the direction of the Orion Nebula. None of the objects proposed earlier as a possible X-ray source - $\theta^2$ Ori A, $\theta^1$ Ori B, BM Ori and $\theta^1$ Ori C - seems to emit X-rays that were tentatively explained as due to the coronae around the T Tauri stars within the nebula \cite{DenBoggende+1978}.

Only a handful of "normal" stars were detected in X-rays till 1979; the evidence of an high temperature emission present in several RS~CVn's indicates the need for some non-gravitational confinement of emitting plasma pointing toward a possible role of the magnetic field \cite{Walter+1980}.

In the '80s the \textit{Einstein} observations in the 0.15-4.0 keV bandpass had shown, for the first time, that the X-ray emission from normal stars was a much rich and complex phenomenon than the mere extrapolation of solar coronal emission \cite{Vaiana+1981ApJ}, largely because the X-ray luminosity, $L_X$, scales with the square of stellar rotational velocity \cite{Pallavicini+1981ApJ} that declines with increasing stellar age. As a result $L_X$ increases, up to a factor few~$\rm \times 10^2$, with the decrease of stellar ages \cite{Micela+1985, Micela+1988}.

Given the limiting sensitivity of a typical \textit{Einstein} observation, $f_X \rm \sim 4-8 \times 10^{-13}$ erg/cm$^2$/sec, assuming as template the Sun with $L_X \rm \sim 1-3 \times 10^{27}$ erg/sec (e.g. \cite{Judge+2003}), the expectation was to detect only a few coronal sources among the about 50-100 stars within 5--8 parsec from the Sun. When all data were carefully analysed and scrutinised more than 1100 normal stars were detected \cite{Sciortino1993}: solar-like stars up to distances of a few hundreds parsec and OB stars up to a few kilo-parsecs. The overall \textit{Einstein} stellar survey includes about 100 YSOs. The \textit{Einstein} observatory stellar results have been extensively reviewed in several occasions \cite{Rosner+1985, Vaiana+1992MmSAI, Sciortino1993}.

A handful of \textit{Einstein} IPC observations of few selected SFRs, like Orion, Taurus-Auriga, Corona Australis, LDN981 in Cygnus, and few more
had shown that a fraction of T-Tauri stars emit X-rays with $L_X$ in the $\rm \sim 1-100 \times 10^{30}$ erg/sec range and had discovered previously unknown YSOs with $L_X \rm \sim 10^{30}$ erg/sec, namely at the attained limiting sensitivity for the source distances and absorptions. Regarding the origin of the X-ray emission several possible alternatives were proposed. For low-mass YSOs ($M \lesssim 2 M_{\astrosun}$) the observed $L_X$ level had left as likely the explanation of an extension of the process at work in late-type main-sequence stars. This process can naturally account for the typical, $\rm \sim few \times 10^6$ K, temperature of emitting plasma and for a factor 10, rapidly rising flare detected in one YSO. 

The extensive \textit{Einstein} survey of the center of the $\rho$ Oph SFR \cite{Montmerle+1983ApJ}, originally planned in search of COS-B counterparts, has definitively opened the avenue of the X-ray studies of SFRs with the detection of about 50 X-ray sources, 2/3 of which had as likely counterpart a known YSO, while 1/3 of X-ray sources had (at that time) no-known IR/optical counterparts. The derived $L_X$ were $\rm \sim 10^2$--$\rm 10^3$ more intense than the solar one and most of the sources were highly variable: up to a factor 10 on a time scale of a day. The distribution of (variable) source normalized amplitude variations followed a power law. The variability was interpreted in terms of strong stellar flares with the implication that the observed emission was, as in the case of solar corona, a manifestation of complex magnetic phenomena. The limited spectral resolution and time coverage of data prevented to draw firm conclusion on the location of the emitting plasma and on the role played by the circumstellar disks. 

Analysing all available \textit{Einstein} IPC data, the nearby Chameleon (Cha) I SFR was studied: 22 X-ray sources were found, 17 associated with known YSOs members, even if source confusion made identification ambiguous in about half of the cases \cite{FeigelsonKriss1989ApJ}. About 2/3 of detected population were Class~III YSOs and 1/3 were Class~II YSOs. The Cha~I likely members resulted to have an average $L_X \rm \sim 10^{30}$ erg/s 3 times higher that the Pleiades dK stars. By comparing the Cha~I, the Pleiades and the Hyades X-ray Luminosity Distribution Functions (XLDFs) the monotonically decrease of $L_X$ with age (cf. \cite{Micela+1988} and references therein) found in the age range $\rm 10^{7.8} - 10^{10}$ yr was extended down to $\rm \sim 10^6$ yr.

\begin{figure}[t]
\vspace{-0.2cm}
\includegraphics[width=6.0cm]{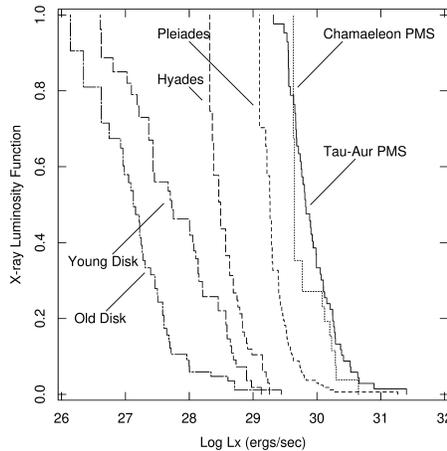}
\texttt{\sidecaption
\caption{The \textit{Einstein} stellar survey XLDFs derived with the Kaplan-Meier estimator from detected and undetected X-ray members of selected open-clusters and SFRs. In Taurus-Auriga and Chameleon the available data does not allow to derive XLDFs separately for Class~III and for Class~II/I YSOs. The $L_X$ of YSOs are on average higher than that of zero-age main-sequence (ZAMS) stars (adapted from \cite{Damiani+1995}); for a recent version from \textit{Chandra} results see Fig. 4 of \cite{PreibischFeigelson2005ApJS}.
}
\label{fig:1}
}
\end{figure}

The systematic, refined analysis of the about 4000 \textit{Einstein} archived observations, had enabled the comprehensive \textit{Einstein} study of the Taurus-Auriga YSO population \cite{Damiani+1995, DamianiMicela1995}. 
The comparison of the derived XLDF with those of other coeval stellar groups had shown that the steady increase in the average level of X-ray emission toward younger ages does extend to the pre-main-sequence phase (see Fig. \ref{fig:1}).  
A multi-wavelength analysis of X-ray and other activity diagnostics provided constraints on the origin of the X-ray emission and on the structure of the YSO outer atmospheres. A significant difference was found between the distributions of L$_X$ of YSOs with and without strong emission lines (i.e. likely Class~II vs Class~III YSOs), but no difference was found between YSOs with and without strong infrared excess at 25 {\textmu}m\footnote{This is one of the first cases in which a different behaviour seems to emerge between the accreting and the disk-bearing YSO samples}. For age younger than 2 $\rm 10^6$ years, anti-correlation between X-ray luminosity and rotational period was found supporting the magnetic origin of the X-ray emission. At later ages saturation of X-ray surface flux was found to take place. In the case of Class~II YSOs (e.g. Classical T-Tauri objects) still bearing a substantial disk, an inverse relation between $L_X$ and the H$\alpha$ emission scaled to the infrared (disk) emission was found and was interpreted as evidence of a competition between closed and open magnetic structures on the stellar surfaces. In the Class~III YSOs, instead L$_X$ was correlated with the unscaled H$\alpha$ luminosity, providing additional evidence for the chromospheric origin of the line emission.
\begin{svgraybox}
At the end of '80s it was clear the key role of X-ray observations in the study of SFRs and the opportunities that the improvements of instrumental capabilities that were either planned or under study would have opened.
\end{svgraybox}
\section{ROSAT and the nearby star forming sites}
\label{sec:3}
In 1990 ROSAT has further advanced the X-ray studies of SFRs.
ROSAT was initially devoted to perform with the Position Sensitive Proportional Counter (PSPC) the All-Sky Survey (RASS), afterward it was dedicated to pointed guest observations with the PSPC and the US-provided High Resolution Imager (HRI).

The typical RASS limiting sensitivity was $f_X \rm \sim 2 \times 10^{-13}$ erg/cm$^2$/sec, the final source list includes about 135.500 sources \cite{Boller+2016}. A substantial fraction of them have as counterpart a normal star or a YSO; considering only the part of the sky with $ \lvert b \rvert > 15^{\circ}$ and excluding the areas near the Large and Small Magellanic clouds, out of about 106500 sources, at least 10\% of them have a galactic object (i.e. with a detectable proper motion) as a counterpart \cite{Salvato+2018}. The RASS X-ray sources associated to stars/YSOs dominate the galactic plane region and overall about 30\%  are stars/YSOs (\cite{Schmitt1999hxra} and references therein).

With the RASS a widespread population of active, X-ray luminous, stars have been found not only within or in proximity of known SFRs, but also rather far-away from them (a detailed account is provided by \cite{Neuhauser1997Sci} and references therein). This was perceived as a major surprise, even if the analysis of the stellar content of the EMSS (Extended Medium Sensitivity Survey) \cite{Gioia+1984EMSS, Fleming+1988ApJ} based on the lithium abundances derived from high-resolution optical spectra \cite{Favata+1993EMSS} and on the modelling of the X-ray stellar content of the Galaxy, \textit{X-Count}\footnote{Based on a modified version of the Bahcall and Soneira thin-disk population model \cite{Bahcall+Soneira1980ApJS}} \cite{Favata+1992Xcount, Sciortino+1995Xcount}, has had already shown the existence of a stellar population with an age of about 100 Myr (Pleiades-like or ZAMS) and a $\sim$ 100 pc vertical scale-height. The Hipparcos distances of the EMSS stars have shown that, in majority, they are main sequence stars \cite{Micela+1997Hipparcos}. \textit{X-Count} predicted their presence in any X-ray survey \cite{Micela+1993Xcount}. The nature of this widespread stellar population has been debated between those that claimed to consist of weak-lined T-Tauri stars (i.e. Class~III YSOs) (e.g. \cite{Krautter+1997AAS}) and others that instead show compelling evidence that most of them should be explained with a ZAMS population, still young, but no so-young as for the case of the putative Class~III YSOs \cite{Favata+1997WLTT, Briceno+1997AJ, Micela+1997Hipparcos}.
 
The analysis of the stellar sample selected by cross-correlating the RASS and the Tycho catalogues, has confirmed the existence of a galactic density gradient from the plane to the pole \cite{Guillout+1998aAandA, Guillout+1998bAandA} predicted by \textit{X-Count} and by another, independently developed, age dependent stellar population model \cite{Guillout+1996AandA}. 

\begin{figure}[t]
\centering
\includegraphics[width=10cm]{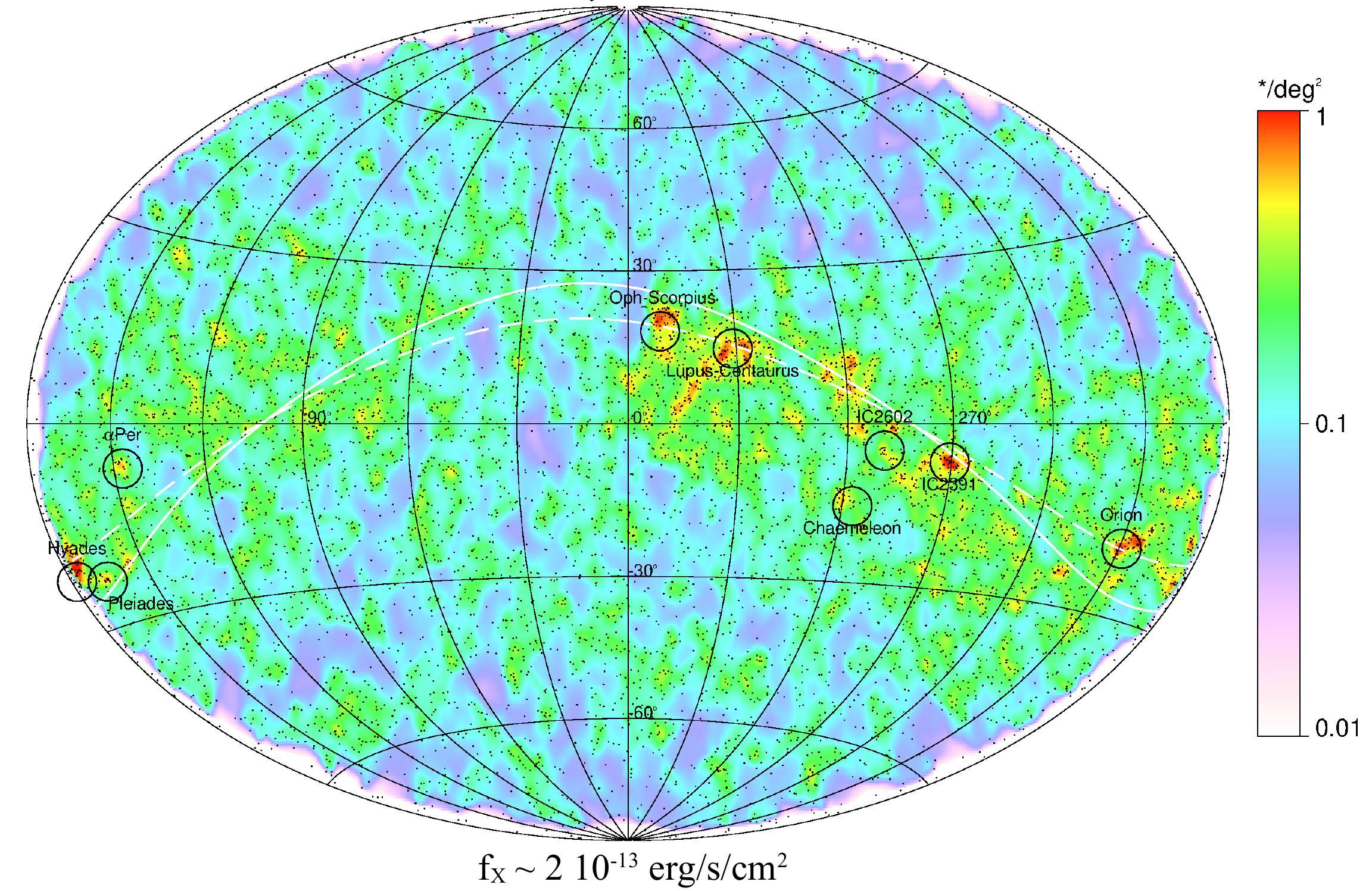}
\caption{Galactic coordinated all sky representation of the 8593 RASS sources (black dots) -- at the limiting f$_X \rm \sim 2~10^{-13}$ erg/s/cm$^{2}$ -- having as counterpart one of the Tycho catalogue star. Note the density enhancement at low galactic latitudes and the asymmetry with respect to the galactic plane.
The dashed line indicates the position of the Gould Belt, while the solid line
is the best fit to the density of matching stars using an exponential flat disk
distribution. Black circles indicate the positions of
the more conspicuous young open clusters and SFRs that show up in the RASS-Tycho sample. Note the enhancement between IC~2602 and 
Lupus-Centaurus toward the Lower Centaurus Crux OB association (adapted from 
\cite{Guillout+1998aAandA}).}
\label{fig:2}
\end{figure}

More surprising was the occurrence of a low galactic latitude feature asymmetric respect to the galactic plane and
located between $l = 15^{\circ}$ and $l = 195^{\circ}$, where the concentration of X-ray active, very young late-type stars (${\rm Log}~L_X$ [erg/sec] = 29.5 -- 30.5) is 
higher than expected. Such an excess was interpreted as due to young late-type (mostly F and G) stars belonging to the so-called Gould belt or 
ring\footnote{It could be a ring like structure of which we recognize only the part nearest to the Sun.}\cite{Guillout+1998aAandA}. This structure encompasses several star formation sites and very likely
hosts a population of bona-fide Class~III YSO, while most of the widespread population consists of near ZAMS stars. The Lithium abundances derived from high resolution spectra provided
convincing evidence that the ROSAT-discovered stars near the Lupus SFR are Class~III YSOs \cite{Wickman+1999MNRAS}. As extensively discussed in this study, the area around the Lupus is very likely the best place to find as clean as possible a population of bona-fide WTTS, while around other nearby SFRs like Chameleon, Taurus-Auriga, etc. the situation is much less favourable and substantial intermixing of ZAMS stars and bona-fide YSOs occurs. 
In this scenario the solar neighbourhood would be located in a region of the Galaxy characterized by a high spatial density of young stars. The origin of the (young) flare stars around the Sun is an old astrophysical problem \cite{Ambazsumian1957} connected to the characteristics of the birth environment of the solar system (cf. \cite{Adams2010review} for a pre-GAIA review). The RASS and follow-up observations provided additional hints for a tentative explanation on the nature of the nearby stellar population.  However the recent GAIA results are challenging the paradigm of the nearby (distance $<$ 500 pc) star formation being part of an ordered belt/ring-like structure (i.e. the Gould Belt) and instead support a scenario where, when a detailed view can be obtained, star formation is complex and filamentary \cite{Zari+2018}. This is 
intrinsically connected to the existence and nature of the so-called moving groups.
The interested reader is referred to the Young moving group chapter of this Handbook section.

The RASS has firmly shown that (wide-field) imaging X-ray observations are extremely efficient in finding ZAMS stars and YSOs from  $\rm \sim 10^5$ to $\rm \sim 10^7$ years, but for the very early evolutive stage since no bona-fide Class~0 YSO has been detected. X-ray observations have shown to be essential to find the Class~III YSOs that lacking a conspicuous circumstellar disk cannot be traced by IR imaging survey and whose identifications, based on high resolution spectroscopy, was (and still is) too demanding. 
The RASS limiting sensitivity and angular resolution permit to follow this powerful approach only for the (not too crowded) SFRs within 200-300 pc. This has provided a large sample of YSOs with known $L_X$ (or upper limits for the undetected known YSOs); most of the known Class~III YSOs and a small fraction of Class~II YSOs were detected (cf. \cite{Neuhauser1997Sci} and references therein).

The ROSAT PSPC bandpass (0.1-2.4 keV) and spectral resolution (0.43 at 0.93 keV) have enabled limited detailed investigations of the physical processes occurring in the YSO magnetosphere and the studies on the origin of the YSO X-ray emission have concentrated in searching correlations between $L_X$ and other YSO parameters. As an example, ROSAT has left unsolved if the smaller fraction of detected Class~I and II YSOs was due to a larger X-ray absorption or to a lower intrinsic $L_X$.

The ROSAT pointed programs have permitted to reach about 10 times deeper sensitivities than the RASS and 
to detect a conspicuous population of low-mass YSOs in
the Chameleon I cloud, in L1495E, and L1551 part of the Taurus complex, in the Cepheus~OB3 region, 
in the Upper Sco-Cen region, around the Orion Trapezium, in the few Myr-old clusters IC348, NGC1333 as well as a population of embedded YSOs in the $\rho$ Ophiuchi and R CrA cloud cores. The X-ray detections of low-mass YSOs have provided the first evidence of a few Myr cluster around $\sigma$ Ori.  

In the Cha~I SFR, a pointed observation has shown that Class~II and Class~III YSOs are coeval and among the
X-ray detected YSOs the Class~III outnumber the Class~II by at least a factor 2 \cite{Feigelson+1993ApJ}. 
$L_X$ resulted to be strongly correlated with mass, effective temperature, bolometric luminosity, and radius; the surface X-ray flux, $F_X$, scaled as $\propto R_*$ implying no magnetic saturation of the stellar surface. However, no firm conclusion was reached on the 
variable(s) that causally determine the X-ray emission level. A "standard" dynamo mechanism cannot reproduce the relationships found, unless the dynamo efficiency drastically changes with the evolution along the Hayashi track. Dynamo model test by exploring the relation between $L_X$ and the rotational period, $P_{Rot}$ or the Rossby number (defined as the ratio of the rotation period to the convective turnover time, \cite{Noyes+1984ApJ}) was prevented for the lack of rotational data and of   Rossby number evaluation in the pre-main sequence phase. For the main-sequence and giant stars the extensive analysis \cite{Pizzolato+2003AandA,Wright+2011ApJ}, has
shown the existence of two emission regimes, one in which the star $P_{Rot}$ is a good predictor of the X-ray luminosity, and the other in which $L_X/L_{Bol}$ saturates at $\rm \sim 10^{-3}$. The rotation periods below which stars enter the saturated regime depend on stellar masses.

In the deep PSPC survey of the central region of the $\rho$ Oph SFR
55 reliable and about 50 candidate X-ray sources have been found, doubling the number of \textit{Einstein} sources \cite{Casanova+1995rhoOph}. 
For 1/3 of the reliable X-ray sources no optical counterparts have been found down to $R \sim$ 18 indicating that they are embedded in the cloud and likely very young. Most of identified X-ray sources had as counterpart a Class~II or a
Class~III YSO; about 10 sources had a Class~I YSOs as a possible counterpart, uncertain because of spatial confusion. Several tens of new SFR members have been found and more than 70\% of the Class~II and Class~III members have been detected.
$L_X$ and $L_{Bol}$ are strongly correlated, as for the likely older Cha~I YSOs: $L_X/L_{Bol} \rm \sim 10^{-4}$. $L_X$ ranges from 10$^{28.5}$ to 10$^{31.5}$ erg/s and the XLDFs of the embedded and visible YSO samples were statistically indistinguishable. The numerous population of X-ray emitting embedded YSOs raised the question of the effect of their X-rays on the local environment. They should produce a partial ionization of the inner regions of the circumstellar disk, possibly coupling magnetic fields and wind or bipolar outflows. X-rays should also play a role as a feedback agent in self-regulating star formation if the YSO X-rays exceed (or are comparable to) the cosmic-rays as the principal source of ionization in molecular cloud cores. In the case of $\rho$ Oph, simple calculation suggested that X-rays indeed dominate, but a more detailed modelling was required (see Sect. \ref{sec:5.6}). 

The determination of $L_X$ of Class~0 and Class~I YSOs is a required step for answering this question. Using archival ROSAT PSPC pointed observations a survey of 156 YSOs comprising low-temperature objects in nearby SFRs, H-H outflow sources, CO outflow sources, and Class~0 YSOs has been performed \cite{Carkner+1998AJ}. Only 11 YSOs have been detected and only 1, Ced 110 IRS 6, 
was a new Class~I YSO bringing to the conclusion that the very X-ray luminous ($L_X \rm > 10^{31}$ ergs/s) protostars are rare. No bona-fide Class~0 YSO was found to emit X-rays.

There have been attempts to explore the effectiveness of X-rays in tracing the low- and intermediate-mass SFRs at distances $>$ 1 kpc \cite{Gregorio-Hetem+1998}. In this study, plausible optical counterparts were found for most of the ROSAT sources, except for about 25\% of them that are probably embedded in the clouds. Near-IR data were consistent with the sources being YSOs with $L_X$ in the $\rm 10^{30}$ -- $\rm 10^{32}$ erg/sec range, like in the nearby SFRs. The detection of individual YSOs
confirmed that X-rays do efficiently trace low- and intermediate-mass star formation at significant
distances across the Galaxy. The limited PSPC angular resolution resulted in source confusion from groups of embedded sources, as it was already the case in the $\rho$ Oph core F. Source confusion was problematic in the center of the most crowded regions, like  the Orion Trapezium \cite{Gagne+1995OrionHRI}, even if the observations were performed with the HRI having an angular resolution of $\sim$ 5".

Last, but not least, both ROSAT and \textit{Einstein} were operating on a low-earth orbit with a period of about 1.5 hours, then data gathering has to stop during each south-Atlantic anomaly passage because of elevated particle background, hence continuous uninterrupted observations were rather short ($\sim$ 1 hour) clearly affecting the studies of YSO variability, a clear signature of their X-ray emission known since the \textit{Einstein} observations. 
An illustrative example is provided by the ROSAT HRI study of the Orion Nebula Cluster region \citep[cf. Fig. 11 in][]{Gagne+1995OrionHRI} where 10 low-mass YSOs
have shown evidence of intense flare-like variability. The short duration of the observation windows prevented to observe in 5 YSOs the flare rise and in 2 more YSOs the end of flare decay and all light curves were affected by
substantial data gaps. In two cases -- P1846 (LY Ori, K7e) and P1977 (AG Ori, G8-K0e) -- the inferred rise times were 8 and 9 ksec and the decay times 35 and 45 ksec, respectively;  these values are at odds with those of solar flares and pose the question of these flare nature. 
The characteristics and nature of the magnetic structures where the YSO flares occur is a question that only the \textit{Chandra} data have permitted to investigate (see Sect. \ref{sec:5.4}).

\section{ASCA: looking for X-rays from Class~I and Class~0 YSOs} 
\label{sec:4}
In 1993 the Japan-US X-ray mission ASCA has provided, for the first time, CCD-resolution ($\sim$ 150 eV) time resolved X-ray spectra of a few nearby YSOs up to the Fe 6.7 keV line and even above. Several ASCA observations have
investigated the onset of the (hard) X-ray emission among Class~0 and Class~I YSOs,
something that was impossible with ROSAT whose sensitivity drops above $\sim$ 2.4 keV. 
However, even in the nearby SFRs, a limiting factor was the angular resolution (2.9 arc-min, HEW)\footnote{This was mitigated by the mirror point spread function shape with a sharp central peak that has allowed to distinguishing two sources with a separation of $\sim$ 0.5 arcmin}.

Several observations have surveyed the core of nearby molecular clouds hosting groups of YSOs in several SFRs: 
R~Cra, L1630, the core F of $\rho$ Oph, Monoceros R2, NGC~2023 and NGC~2024 and two cores in the Perseus complex. 
About ten point-like sources have been typically detected in each observation, most of them have had as counterpart one or more YSOs of Class~I or II and more rarely of Class~III.
Several Class~0 YSOs were observed and, notwithstanding the amount of devoted observing time, no bona-fide Class~0 YSO was detected.

In R~CrA hard X-ray emission from Class~I YSOs has been found for the first time \cite{Koyama+1996PASJ}. The quiescent X-ray spectrum was modelled as an absorbed ($N_H \rm = 4.2 \times 10^{22}$ cm$\rm^{-2}$) thermal bremsstrahlung (kT = 7.2 keV) plus a line emission at 6.45 keV. In $\rho$ Oph during a flare occurred on the Class~I YSO Elias~29 the spectrum has shown a factor 10 increase in the absorbing column (N$_H$), consistent with a scenario in which the flare emission comes from a more "external" region than the one emitting the quiescent emission and is affected by additional absorption due to an anisotropic distribution of circumstellar matter, like in the case of a circumstellar disk. In YLW~15, a class~I YSO in $\rho$ Oph previously detected with the ROSAT HRI \cite{Grosso+1997Nature}, 3 subsequent flares occurred every $\sim$ 20 hours have been detected. The last two flares were consistent with the reheating of the same magnetic structure hosting the first flare due to the interaction between the star and the disk as a results of the differential rotation \cite{Tsuboi+2000ApJ,Montmerle+2000ApJ}. 

The ASCA observations have shown that the X-ray emission
mechanism from Class~I and II YSOs during the quiescent state requires 
plasmas with kT $\gtrapprox$ 6 keV, but keeping almost the same 
$L_X /L_{Bol}$ ratios typical of Class~III YSOs. The enhanced 
disk-magnetosphere interaction model \cite{Shu+1997Science} was suggested to account for such
high-temperature plasmas. 
The observed hard X-ray flares were consistent with a model based
on a magnetic loop connecting the central star and its disk \cite{Hayashi+1996ApJ}.
Still today their existence is an intriguing, and controversial, issue (see Sect. \ref{sec:5.4}). 

\section{The transformational impact of \textit{Chandra} and \textit{XMM-Newton}}
\label{sec:5}

The boost of observational capabilities enabled by \textit{Chandra} and \textit{XMM-Newton} 
has made it possible to perform up to $\sim$ 150 ksec long uninterrupted high sensitivity observations with energy resolution of $\sim$ 100 eV, angular resolution of 0.5-10 arc-sec over field-of-view (FOV) ranging from $\sim$ 0.08 to 0.18 sq. deg.  Time-resolved X-ray CCD spectroscopy of YSOs in SFRs up to a distance of a few kpcs has become possible and, together with
high-resolution X-ray spectroscopy of a handful of nearby YSOs, have had a profound impact on our knowledge. 
Thanks to \textit{Chandra} superb PSF the sensitivity and angular resolution has allowed us to survey SFRs up to distances of $\sim$ 4 kpc, even in their denser subgroups. These data have given rise to transformational investigations along two major avenues: 1) the study of the processes originating the YSO X-ray emission and their changes with evolution, something that was started before but mostly left unsolved; 2) the study of the complex processes that result in star cluster formation, especially of those hosting OB stars, that was essentially impossible before. The former has been tackled already with the very early observations
and has been mostly based on the SFRs within $\sim$ 1 kpc, while the second has emerged more vigorously when enough data have been available and uniformly analysed and has been based on a much larger number of SFRs up to $\sim$ 4 kpc. A crucial role has been played by optical/IR/NIR observations that have enabled a successful multi-wavelength approach.  

The shape of the star cluster mass function, $\rm{d} N/\rm{d} M \sim M^{-2}$, implies that $\rm >50\%$ of stars form within massive SFRs containing OB stars where rich clusters born, likely constituting the main mode of star formation in the Galaxy. In the early 2000s several aspects of the stellar cluster formation were poorly known or controversial, as discussed in various reviews and papers (\cite{Lada+Lada2003ARAA,Allen+2007prpl.conf,Kennicutt+Evans2012ARAA,Motte+2018ARAA,Feigelson+2013ApJS} and references therein). Let me recap a few of them:
i) the effect of the new born OB stars on cloud ionization and dispersal \cite{Bate2009MNRAS,Howard+2016MNRAS,Rumble+2021MNRAS}; 
ii) the duration of star formation with the two alternative scenarios of a very rapid star formation \cite{Elmegreen2000ApJ}, and of a star formation process progressing actively for millions of years \cite{Tan+2006ApJ}; 
iii) the issue of rich cluster formation as a global process or as the result of the merging of smaller groups \cite{McMillan+2007ApJ, Maschberger+2010MNRAS}; 
iv) the origin of the age spread often found in the Hertzsprung-Russel diagram \cite{Baraffe+2009ApJ,Hosokawa+2011ApJ,Jeffries+2011MNRAS,Jensen+2018MNRAS,Prisinzano+2019AA} and its relation to the duration of star formation process; 
v) the relevance of HII region expansion in triggering star formation (e.g. \cite{Ogura+2007PASJ,Getman+2009ApJ}); 
vi) the most relevant mechanism of massive star formation among the many proposed: monolithic collapse, stellar mergers, rapid disk accretion, and competitive accretion (cf. \cite{Motte+2018ARAA} and references therein); 
 vii) the effect on proto-planetary disk survival of the massive star irradiation \cite{Johnstone+1998ApJ,Ercolano+2008ApJ,GortiHollenbach2009ApJ} and/or stellar density (e.g. \cite{ClarkePringle1993MNRAS, Thies+2010ApJ}) within rich clusters. 

For decades cluster formation studies have been hampered by the observational difficulty to obtain a reliable and unbiased member census. At low galactic latitude, in the optical,
(older) field stars have a surface density 10-100 times higher that SFR members at the peak of their IMF, A$_V$ can be as high as 30 and can vary by tens of magnitudes within a given SFR. YSOs are difficult to detect as faint infrared objects toward the background emission due to heated dust in HII nebulae. As a result often only the SFR bright cores, where stellar density is very high, have been identified, the OB stars have been recognized by their color and affordable spectroscopic studies; instead low-mass YSOs have been identified by their photometric infrared excess due to the circumstellar disk. This procedure recognizes the disk-bearing YSO members, but fails to select the, usually, larger population of YSO members that have already lost (most of) their circumstellar disk. This latter population is "easily" selected with medium-deep imaging X-ray observations because the low-mass YSO X-ray luminosity is at least 3 dex higher than the low-mass field stars\footnote{The contaminant AGNs seen through the galactic plane can be discarded since lack the infrared counterparts}. In a 100 ks \textit{Chandra} observation of a SFR at a distance of 2-3 kpc about 1-2 10$^3$ YSO members are typically detected allowing to sample the IMF typically down to $\sim$ 0.5 $M_{\astrosun}$ (e.g. \cite{Damiani+2004ApJ}) and even to BDs in the nearby SFRs (e.g. \cite{Imanishi+2003PASJ}) or farther away in deeper observations (e.g. \cite{Preibisch+2005ApJSCOUP_BD}). With a judicious combination of X-ray and IR selected YSO members recent and on-going studies based on almost unbiased samples have been possible. 

New avenues to investigate some of the questions on cluster formation process have recently been opened by GAIA \cite{Brown2021ARAA}. The quality of DR2 and EDR3 GAIA data allows, within $\sim$ 1 kpc, detailed studies of the low mass, down to $\sim$ 0.1 $M_{\astrosun}$, star population of the young clusters and stellar associations. 
On the other hand, given the GAIA bandpass, the very young, more absorbed and/or more embedded, stellar population is much more difficult to recognize and characterize (cf. \cite{Prisinzano+2022astroph}).
As an example, in the case of NGC~2264 with the EDR3 data about 50\% of the very likely members selected/confirmed by X-ray and IR data are recovered as members and a sizeable fraction of the embedded ones are missed (\cite{Prisinzano+2022astroph} as well as an ongoing analysis by E. Flaccomio). This effect is expected to become more consistent as the young cluster/SFR distance increases. In summary, while GAIA is adding and will add superb pieces of information on the study of cluster formation still X-ray and IR surveys will continue to play a relevant role 
to build as much as possible an unbiased low-mass member list of young clusters/SFRs beyond $\sim$ 0.5 -- 1 kpc. 

A first round of \textit{Chandra} studies, usually based on 50-100 ksec long ACIS observations, has concentrated on the Orion Nebula Cluster (ONC) region \cite{Garmire+2000AJ,Feigelson+2002ApJ,Feigelson+2003Orion,Flaccomio+2003aApJ,Flaccomio+2003bApJ}, Orion outer regions \cite{Ramirez+2004bAJ}, NGC~1333 \cite{Getman+2002ApJ}, RCW~38 \cite{Wolk+2002ApJ}, W3 \cite{Hofner+2002ApJ}, OMC~2 and OMC~3 \cite{Tsuboi+2001ApJ,Tsujimoto+2002ApJ}, the $\rho$ Oph core \cite{Imanishi+2001ApJ,Imanishi+2003PASJ}, Mon~R2 \cite{Kohno+2002ApJ,Nakajima+2003PASJ}, Sgr~B2 \cite{Takagi+2002ApJ}, NGC~2264 \cite{Ramirez+2004aAJ,Flaccomio+2006AA, Rebull+2006AJ}, 
M17 and Rosette \cite{Townsley+2003ApJ}, Lagoon (NGC~6530) \cite{Damiani+2004ApJ,Damiani+2006AA}, Trifid \cite{Rho+2004ApJ}, NGC~2068 \cite{Grosso+2004AA}, Sh~2-106 \cite{Giardino+2004AA}. In the same years deserve attention the \textit{XMM-Newton/EPIC} studies of nearby SFRs or young associations such as L1551 \cite{Favata+2003AA}, NGC~1333 \cite{Preibisch2003bAA}, Serpens \cite{Preibisch2003aAA}, Chameleon~I \cite{Stelzer+2004AA, Telleschi+2006ESASP, Robrade+2007AA}, \rm{$ \rho$} Ophiuchi \cite{Ozawa+2005AA} and Upper Sco \cite{Argiroffi+2006AA}.

In a 100ks ACIS observation toward the \rm{$\rho$} Oph central region $\sim$ 100 sources with $L_X > 10^{28}$ erg/s have been found, about 65\% have optical/IR counterparts with a substantial number of Class~I to Class~III YSOs and few brown dwarfs \cite{Imanishi+2001ApJ}. About 70\% of Class~I YSOs have been detected and about 40\% of the brown dwarfs have been detected with $\rm{Log}~(L_X/L_{Bol}) \sim$ -5 -- -3, similar to the values of main-sequence stars. In YLW~16A, a class~I YSO, for the first time,  a neutral Fe fluorescent line at 6.4 keV it has been firmly detected. The line equivalent-width (EW) requires an origin from circumstellar gas distributed with a non-spherical geometry as for a face-on circumstellar disk. Combining the data from a later ACIS observation of the same region a total of 195 sources with 71 X-ray flares have been found in YSOs and brown dwarfs \cite{Imanishi+2003PASJ}. Most of them have the typical solar flare shape with fast rise and slow decay, except a few bright flares with unusually long rise phase. By modelling the spectra and light-curves, the time-averaged temperature
($kT$), luminosity ($L_{X}$), rise and decay timescales ($\tau_{rise}$ and $\tau_{decay}$) have
been derived showing that class~I-II YSOs have usually high $kT$, in some case up to 5 keV; the
${\rm Log}~L_{X}$~[erg/s] distributions during flares span the 29.5 -- 31.5 range for all YSO classes,
with a marginal evidence of higher $L_{X}$ among class~I YSOs; positive and negative log-linear correlations have been found between $\tau_{rise}$ and $\tau_{decay}$, and $kT$ and $\tau_{rise}$. Assuming that the emission is due to magnetic reconnection within a magnetic loop in presence of heat conduction and chromospheric evaporation those correlations are, in most of the cases, consistent with nearly the same loop length, of the order of the central star 
radius ($\rm 10^{10}$ -- $\rm 10^{11}$ cm), for all YSO classes, regardless of the existence of an accretion disk. However a few flares on ROX~31 and YLW~16A with larger rise timescale ($\rm \sim 10^4$ sec), requires longer loops with length up to $\rm 10^{12}$ cm (i.e. of the same size of the co-rotation radius).

Thanks to a 100~ksec long ACIS observation toward NGC~2264
a total of 420 X-ray point-like sources have been detected, 85\% with optical and NIR catalogued counterparts. Given their $L_X$ values, more than 90\% of identified X-ray sources are NGC~2264 members thereby significantly increasing the known low-mass member population by about 100 YSOs \cite{Flaccomio+2006AA}.  About 50\% of X-ray sources without counterparts are likely associated with members, most of which previously unknown (obscured) YSOs. 
X-ray activity has been investigated as a function of stellar and circumstellar characteristics by correlating the X-ray luminosities, temperatures, and absorptions with published optical and NIR data. This analysis has confirmed previous findings: L$_X$ is related to stellar mass, although with a large scatter; except during some flares, ${\rm Log}~(L_X/L_{Bol})$ is close to, but always below, the -3 saturation level. A comparison between Class~I-II and Class~III YSOs shows several differences: the former have, at any given mass, activity levels that are both lower and more scattered than the latter; emission from Class~I-II may also be more time variable and is on average slightly harder than for Class~III. In some Class~I-II YSOs there is evidence of extremely cool, $\sim$ 0.1-0.2 keV, plasma which is consistent with being heated by accretion shocks and emitted by plasma at higher density, $n_e \rm \sim 10^{11}$ -- $\rm 10^{13} cm^{-3}$ , than the coronal one as it has been found in TW~Hya \cite{Kastner+2002ApJ} and in other YSOs. This emission component has been recently reviewed by \cite{Argiroffi2019AN}. 

\begin{figure}
\centering
{
\includegraphics[width=5.4cm]  {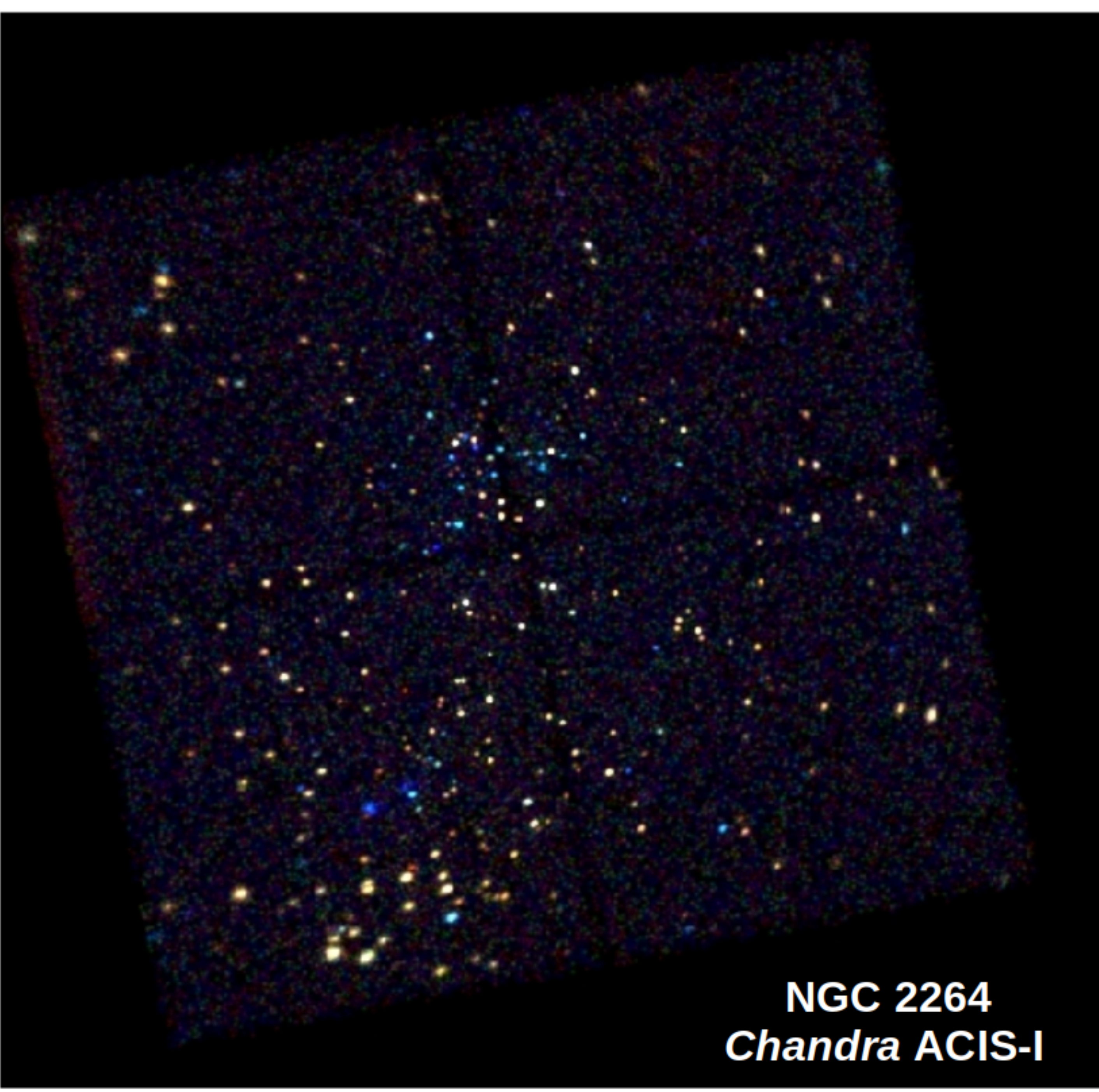}
\includegraphics[width=5.6cm ] {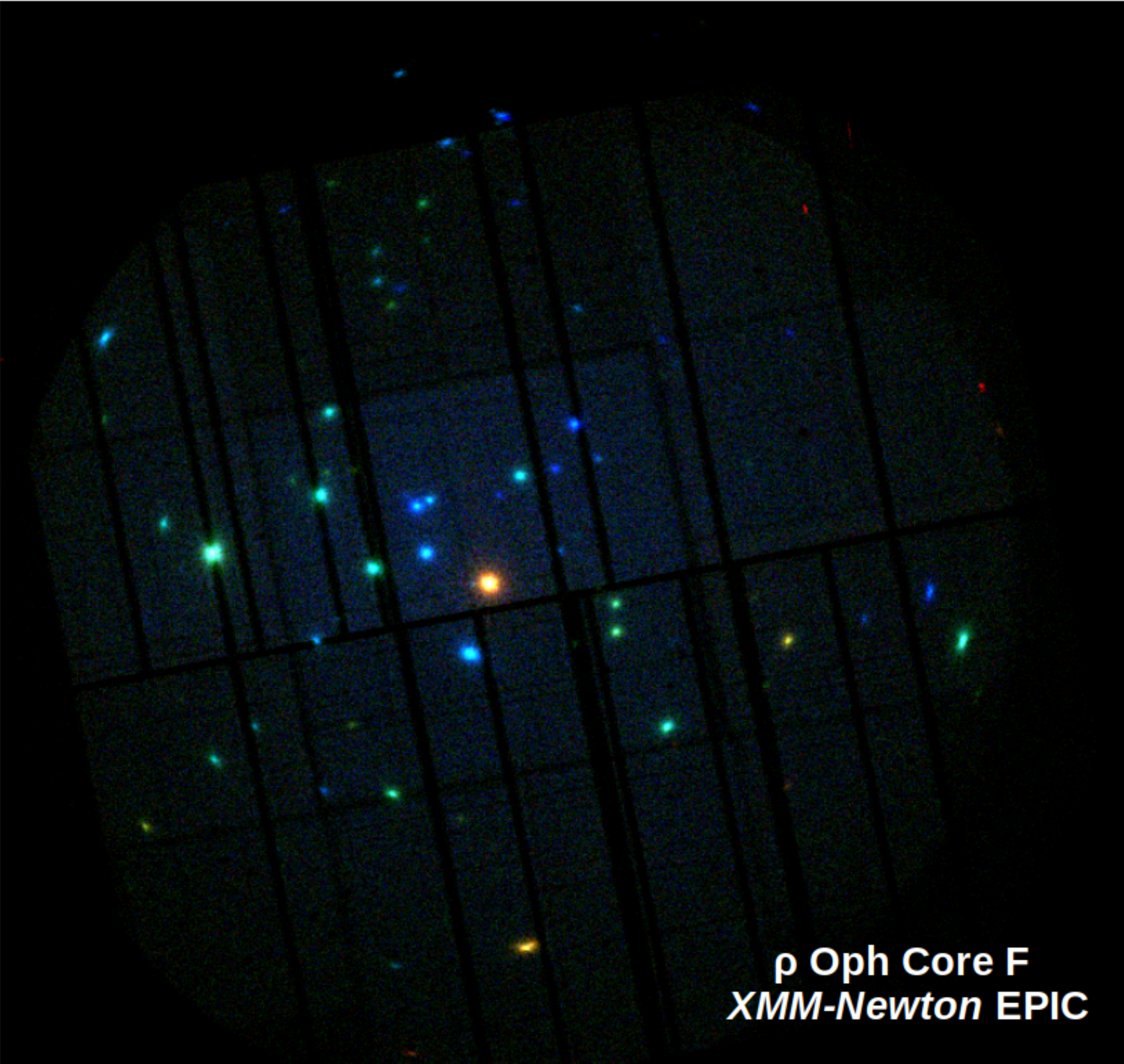}
}
\caption{(Left ACIS-view of NGC 2264. The RGB image is obtained from data in three energy bands: [200:1150] eV (red), [1150:1900] (green), and [1900:7000] (blue). Soft and unabsorbed sources are shown in red, while hard and/or absorbed sources in blue.
(Right) {\protect\textit{XMM-Newton}-EPIC false-color image toward the core F of the $\rho$ Oph SFR. The image is obtained by summing together the 500 ksec DROXO and 350 ks Elias-29 joint NuStar-XMM-Newton Large projects data. High background time interval have been screened out.}}
\label{fig:3}
\end{figure}

The very young cluster NGC 6530 in the Lagoon nebula has been the target of a 60 ks long ACIS observation; 884 X-ray point sources have been detected, 90\%-95\% are likely members, mostly low masses YSOs resulting in a substantial increase of the NGC 6530 YSO population respect to previous optical and H$_{\alpha}$ surveys \cite{Damiani+2004ApJ, Damiani+2006ApJ}. Only $\sim$ 25\% of the X-ray sources have a counterpart down to V=17, mostly have fainter counterparts, 2MASS IR counterparts have been found for $\sim$ 83\% of the X-ray sources. The H-R diagram of the optical counterparts shows that they are above the main sequence and fall in the locus of 0.5--1.5 Myr YSOs, with masses down to 0.5--1.5 $\rm M_{\astrosun}$. 
An age gradient from north-west to south is evident, and is qualitatively consistent with a sequence of star formation events reported in earlier studies. By combining the \textit{Chandra}, optical and 2MASS data and considering various different indicators of IR excesses and reddening-free indices, 333 YSOs with optical-IR excess have been found, 196 in the ACIS fov, 76 undetected in X-rays.  The total number of estimated cluster members thus has become $\geq$ 1100. The estimated disk frequency in the ACIS field was ~20\%. YSOs with optical-IR excess in the north of NGC 6530 are nearly 
co-spatial with a sub-population of older members than the ones in the cluster center. Hence in these northern regions, far from massive cluster stars, star formation and disk evolution has proceeded almost undisturbed for long time while near the cluster center, where most massive stars are found, most of the members lack substantial disks and strong accretion.

The Serpens dark cloud at a distance of $\sim$ 260 pc was studied with a deep \textit{XMM-Newton/EPIC} observation detecting 45 individual X-ray sources. Among their counterparts only one Class~I, and two flat-spectrum YSOs have been found, but none of the Class~0 YSOs was detected \cite{Preibisch2003bAA}. The core of the Serpens cloud was studied also with a 90 ksec \textit{Chandra} ACIS observations with the specific aim to search for the X-ray emission from the 6 known bona-fide Class~0 YSOs falling in the ACIS FOV \cite{Giardino+2007aAA}. None of them was detected, their X-ray emission has remained undetected even by co-adding the data at the 6 individual Class~0 positions (reaching an equivalent observing time of $\sim$ 540 ksec); the derived upper limit to the 
Class~0 source X-ray emission was $L_X <$ 4~10$^{29}$ erg/sec (for $N_H$= 4~10$^{23}$ cm$^{-2}$ and kT = 2.3 keV). Such a value is below the typical \emph{persistent} X-ray emission of the other YSOs in the region. Hence either Class~0 X-ray emission is intrinsically much weaker or they are hidden
behind an absorbing column density substantially higher than 4~$\times$~10$^{23}$~cm$^{-2}$, and/or, admittedly less likely, their X-ray emission is extremely variable.

In the ONC ACIS-based study more than 1000 sources have been found, 91\% of them were known members ranging 
from brown dwarfs to O stars \cite{Garmire+2000AJ,Feigelson+2002ApJ,Feigelson+2003Orion}. About 7\% of X-ray sources were newly identified deeply embedded cloud members. For ONC sources ${\rm log}~L_{X,[0.5-8.0 kev]} {\rm [erg/s]}$ ranges from the sensitivity limit of $\sim 28 $ to $33.3$, and absorption ranges from ${\rm Log}~N_H {\rm [cm^{-2}] < 20.0}$ to $\sim 23.5$. Thanks to the availability of bolometric luminosities, masses, ages, disk indicators and rotational periods it was concluded that: the presence of rapid variability in $\rm \Theta^2$ A Ori, a O9.5, 31 M$_{\astrosun}$, star, and in several early B stars, is at odd with a mechanism of X-ray production due to shocks distributed throughout the radiatively accelerated wind; the low-mass YSOs exhibit large flares but usually $L_X/L_{Bol}$ remains well below the saturation level; the YSO plasma temperature is often very high with T $>$ 100 MK, irrespective of luminosity level and flare presence; a large number of very low-mass objects showing flares with intensity like for the $\sim$ 1 M$_{\astrosun}$ YSOs was found, as well as evidence of the decline of the magnetic activity among L and T type brown dwarfs; $L_X$ is strongly correlated with $L_{Bol}$, the average Log~$(L_X/L_{Bol})$ value is -3.8 for YSOs with masses 0.7 < M < 2 M$_{\astrosun}$, about one order of magnitude below the main-sequence saturation level; $L_X/L_{Bol}$ drops rapidly below this value in some YSOs with $2 < M < 3 M_{\astrosun}$ with no compelling evidence that the intermediate-mass (mid-B -- A type) YSOs are by themselves significant X-ray emitters; $L_X$ shows, if any, a slight increase with rotational periods, in contrast to the strong $L_X$ decline with increasing period seen in main-sequence stars. 
This very different behaviour indicates that the YSO mechanism of magnetic field generation is different from that operating in main-sequence stars (e.g. an $\alpha-\Omega$ dynamo). One of the most promising possibility was, and still is, the so-called turbulent dynamo distributed throughout the deep convection zone, but other proposed models are possible as well. 

The \textit{Chandra} HRC-based study of ONC \cite{Flaccomio+2003aApJ,Flaccomio+2003bApJ} has found, for the first time, convincing evidence of a dependence of $L_{X}$ and $L_X/L_{Bol}$ on circumstellar accretion indicators for which three tentative explanations have been advanced: i) accreting YSOs are less active since rotate slowly because of disk braking, while those accreting less can "freely" spin up and thus saturate their activity. Since no dependence on rotational period nor on Rossby number
has emerged this hypothesis lacks support unless assuming a strong bias on rotational data. 
ii) accretion and/or the presence of a disk and/or out-flows decrease the fraction of the stellar surface covered with the closed magnetic structures from which X-ray emission originates. A similar outcome would result from "non-standard" geometry, as in the case of coronal structures extending to the inner part of disk (e.g. \cite{Montmerle+2000ApJ}).
Because of the inhomogeneous and time-variable nature of accretion, variability studies and simultaneous X-ray/optical observations could help clarify this matter.
iii) Accreting stars have higher X-ray extinction than assumed, so the difference in inferred $L_X$ and $L_X/L_{Bol}$ is only apparent. This could be possible either if the assumed A$_V$-N$_H$ relation, i.e. the gas-to-dust ratio, is different for accreting stars with respect to the average interstellar value or in a scenario in which accretion gas columns that cross the line of sight would obscure X-rays and let optical/IR radiation through. This hypothesis can be and has been tested (see Sect. \ref{sec:5.2.3} and \ref{sec:5.2.4}) with longer exposures that have allowed to confirm that the observed difference is real.

The case of Orion and of other early observations made it clear the need for long continuous observations as well as for simultaneous multi-wavelength observations. The early results have stimulated a large number of guest observing programs that have covered tens of different SFRs up to distances of 4 kpc investigating a range of ages, environmental conditions, metallicities, massive vs. non-massive SFRs, etc. Thanks to superb quality X-ray data about 50 distinct SFRs have been studied often jointly to optical/NIR/MIR archive or purposely taken observations, with a key role played by the IRAC/Spitzer observations that in some cases have been part of coordinated programs. A somehow outdated, but still useful, list of \textit{Chandra} studies can be found in Table~2 of \cite{Feigelson+2013ApJS} and Table~1 of \cite{Getman+2017ApJS}.
 
\subsection{Systematic studies of the star cluster formation process}
\label{sec:5.1}
The very rich \textit{Chandra/ACIS} and \textit{Spitzer/IRAC} archives have made possible two projects: MYStIX (Massive Young- Star-forming complex Study in Infrared and X-ray) aimed to characterize 20 OB-dominated young clusters and their environments within a distance $\leq$ 4 kp (cf. \cite{Feigelson+2013ApJS} and references therein) and SFiNCS (Star Formation in Nearby Clouds), based on MyStIX heritage, aimed at providing a detailed study of the young cluster stellar populations and cluster formation in the nearby (0.2<d<1 kpc) 22 SFRs to be compared with MYStIX richer, more distant clusters (\cite{Getman+2017ApJS} and references therein). Both projects have been based on a homogeneous re-analyses of the \textit{Chandra/ACIS}, the \textit{IRAC/Spitzer} and the United Kingdom InfraRed Telescope observations in order to construct catalogues of SFR members with well defined criteria and map the nebular (hot) gas and dust. The adopted sophisticated analysis has pushed to the very limit the available observations; a thorough discussion of the advantages and possible limitations of produced catalogues is provided by the team (cf. Appendix B of \cite{Feigelson+2013ApJS}). The motivations and results of the MYStIX project have been recently reviewed in detail \cite{Feigelson2018}. 

The MYStIX project has produced a catalogue of 31784 probable SFR members. The analysis of this catalogue has shown that the age spreads within clusters are real with the cluster core formed after the cluster halo \cite{Getman+2014bApJ}. 
There is evidence of older dispersed populations and cluster expansion, and of long-lived, non synchronous, star formation, and, 
at the same time, there is not conclusive evidence of subcluster merging \cite{Kuhn+2015ApJ}. To investigate the spatio-temporal history of the star formation within the studied SFRs a reliable age estimation of various stellar subgroups was needed. The  MYStIX team has developed a simple, but effective estimator \cite{Getman+2014aApJ} based on the well established, but poorly physically understood, empirical correlation between $L_X$ and YSO
mass, $M$, (cf. \cite{Telleschi+2007bAA}) that accounts for most of the 4 decades $L_X$ range found in SFRs. In a nutshell, individual $L_X$, corrected for absorption, provides an estimate of M, while the de-reddened J magnitude provides a proxy for $L_{Bol}$. Using standard evolutionary tracks, $M$ and $L_{Bol}$ provide a (crude) age estimator for each YSO (nicknamed $Age_{JX}$). While individual values are rather inaccurate, median values for spatial defined subgroups appear adequate enough to follow the star formation history within and between SFRs/young clusters.  

The SFiNCs project has built a catalogue of $\sim$ 8500 likely members
increasing by 40\% the census of the 22 analysed nearby SFR regions. The comparative analysis of the SFiNCs and MYStIX  clusters has shown that the former are typically smaller, younger and more heavily obscured than the latter. The SFiNCs clusters associated to molecular clouds have an elongated shape whose major axis is aligned with their host molecular filaments, a firm of the morphology imprint of their parental clouds. Cluster expansion is evident from many indicators.
Core radii increase by one order of magnitude (from $\sim$ 0.08 to $\sim$ 0.9 pc) over the age range 1--3.5 Myr, implying that gas removal time-scale is longer than 1 Myr. There is evidence of an early generation of star formation that has left conspicuous, spatially distributed, stellar populations.

Another project, based on the reanalysis of archival \textit{Chandra} ACIS observations, is the Massive Star-forming Regions (MSFRs) Omnibus X-ray Catalog (MOXC) \cite{Townsley+2019}, that includes X-ray point sources from a selection of 12 MSFRs across the Galaxy,
with distances ranging from 1.7 kpc to 50 kpc, plus 30~Doradus in the Large Magellanic Cloud. MOXC reports 20,623 X-ray point sources. Taking advantage of the point source catalogue, and removing their contributions it has been possible to study the morphology of the remaining diffuse X-ray emission that trace the bubbles, ionization fronts, and photon-dominated regions that are found in all MOXC MSFRs. As already found in the cases of M17 and Rosetta this unresolved X-ray emission is dominated by hot plasma from massive star wind shocks \cite{Townsley+2003ApJ}. This diffuse X-ray emission shows that massive star feedback (and the accompanying several-million-degree plasmas) is a key element of MSFR physics. 

\subsection{Long-look, large area and multi-wavelength simultaneous surveys}
\label{sec:5.2}
In the current scenario of Class~I-II YSOs \cite{Hartmann2008book} magnetically funnelled accretion streams connect the central star with its circumstellar disk. In such a system X-rays could be emitted by the PMS star corona, by the funnel plasma that is shocked as it accretes on the star, by the fluorescing disk matter or by gas shocked in a jet. X-ray studies allows us to disentangle those contributions and to investigate the processes at work. Among the many studies particularly notable are the long look programs and the large-area surveys that have consolidate emerging scenarios or opened new perspectives. 

\subsubsection{NGC~1893: exploring star formation in the outer Galaxy}
\label{sec:5.2.1}
The young cluster NGC~1893 in the external region of the Galaxy at about 12 kpc from the galactic center has been observed with Chandra for 450 ksec. This program, lead by G. Micela, has allowed to study the IMF in a region of the Galaxy where the environmental conditions are different than in the vicinity of the Sun. 
A catalogue extending from X-rays to NIR has allowed to derive the cluster membership as well as other cluster parameters: 415 disk-less candidate members and 1061 disk bearing candidate members, 125 of which are also $\rm H_{\alpha}$ emitters, have been found showing that NGC 1893 contains a conspicuous population of YSOs, together with the well-studied main sequence cluster population. The fraction of disk bearing members is about 70\% as found in similar age clusters in the solar vicinity. Hence, despite expected unfavourable conditions for star formation, evidence has been found that very rich young clusters can also form in the outer regions of our Galaxy \cite{Prisinzano+2011AA}.

\subsubsection{DROXO and follow-on: the enigmatic variability of YSO Fe 6.4 keV line}
\label{sec:5.2.2}
Two long look programs, lead by S. Sciortino, have been devoted to study the $\rho$ Oph core F region. The first is nicknamed {\em DROXO} ({\em D}eep {\em R}ho {\em O}phiuchi {\em
X}MM-Newton {\em O}bservation) and consists of a 500 ks long continuous \textit{EPIC} observation \cite{Sciortino+2006ESASP}. The {\em DROXO} time resolved spectroscopy has allowed us to study the X-ray emission of the 1 Myr old $\rho$ Oph YSOs. A successive joint \textit{XMM-Newton -- NuStar} has been devoted to investigate the existence (if any) and origin of the very-hard X-ray emission and to further perform time-resolved X-ray spectroscopy of the $\rho$ Oph YSO Elias~29, specifically of its Fe 6.4 keV K$_{\alpha}$ line. This line EW had shown intense variability, not clearly associated with YSO flares \cite{Giardino+2007bAA} and tantalizing evidence of line centroid displacement that could either be due to Doppler shift of emitting matter or rapid changes of its ionization \cite{Pillitteri+2019}.
Since its first firm discovery in YLW~16A the Fe 6.4 keV K$_{\alpha}$ has been now found in several tens of YSO 
(e.g. \cite{Tsujimoto+2005COUP, Favata+2005b, Stelzer+2011}) but only in few YSOs observed with DROXO line variability has been investigated in some detail \cite{Giardino+2007bAA, Stelzer+2011, Pillitteri+2019}. 

\subsubsection{XEST and the origin of YSO mass-L$_X$ and accretion-L$_X$ relations}
\label{sec:5.2.3}
Another program devoted to the study of YSO physics was {\em XEST} ({\em X}MM-Newton {\em E}xtended {\em S}urvey of {\em T}aurus \cite{Guedel+2007XEST} and references therein),lead by M. Guedel, based on a mosaic of many medium-deep exposures to cover most of the Taurus region.
The XEST has confirmed the existence of an $L_X$ vs. mass relation that has been interpreted as a consequence of the X-ray saturation and of the mass vs. bolometric luminosity ($L_{Bol}$) relation occurring among YSOs in a narrow age range \cite{Telleschi+2007bAA}. In the saturation regime $L_X \sim \rm{C} \times L_{Bol}$ with Log(C) = -3.73 for Class II and -3.39 for Class~III YSOs, confirming the findings of the intrinsic factor $\sim$ 2 lower $L_X$ of Class~II respect to Class~III YSOs. As a possible explanation the same authors have suggested that some of the accreting material in Class-II is cooling active regions preventing them from contributing to the X-ray emission. An alternative scenario assumes that YSO X-rays modulated the accretion flow by the X-rays heating the gas in the surface layer of the disk (e.g. \cite{Drake+2009ApJ}). If the local temperature excesses the escape temperature a thermal driven wind will start and the accretion rate will decrease: in a few Myr a gap will develop in the inner disk, the erosion will start and the accretion rate will drop. The YSOs with higher $L_X$ will reach sooner the low-accretion rate condition.  Detailed models \cite{Picogna+2019MNRAS, Picogna+2021MNRAS} predict a weak anti-correlation between $L_X$ and the mass accretion rates (see Sect. \ref{sec:5.5}).

\subsubsection{COUP: L$_X$ vs. rotation and age, insights on the dynamo and the origin of saturation}
\label{sec:5.2.4}
The Orion central region has been extensively studied with {\em COUP} (\protect{{\em C}handra} {\em O}rion {\em U}ltradeep {\em P}roject, lead by E. Feigelson, cf. \cite{Getman+2005COUP}), 10 days (850 ksec) of ACIS integration during a continuous time span of 13 days, yielding a total of 1616 detected sources in the 17'×17' field of view which has allowed us to study the X-ray properties of known Orion YSOs as well as to discover and characterize the Orion YSO embedded population. 

The analysis of COUP data \cite{Preibisch+2005ApJSCOUP_Origin} has confirmed with a high confidence level that the L$_X$ level of non-accreting YSOs is consistent with that of rapidly rotating main-sequence stars, while the accreting YSOs have, a factor 2--3 on average, weaker $L_X$. This weakening is not due to the absorption since $L_X$ has been properly absorption-corrected, nor to possible disk-star rotational locking given the lack of correlation of $L_X$ with rotation. The other possible explanations are i) the accretion effects on YSO magneto-sphere and the loading by the accreting matter on part of field lines resulting in higher density, reduced temperature and weaker $L_X$, or ii) modification of the internal stellar structure due to the accretion that weaken the dynamo action. The COUP data have also confirmed the
strong, almost linear, correlation between $L_X$, $L_{Bol}$ and YSO mass already reported in the early studies \cite{Feigelson+2003Orion, Flaccomio+2003bApJ}. $L_X/L_{Bol}$ increases slowly with mass over the 0.1-2 $\rm M_{\astrosun}$ range.
The scatter about these relations is larger than explainable in terms of data uncertainties, unresolved binaries and intrinsic X-ray variability, and it likely due to the effect of accretion on $L_X$. 
Taking advantage of the exceptional COUP time coverage, rotational modulation of X-ray emission was searched on a subsample of 233 (out of 1616 sources) X-ray-bright stars with known rotational periods and was found in at least 23 YSOs with periods between 2 and 12 days and relative amplitudes ranging from 20\% to 70\%. In 16 cases, the X-ray modulation period is similar to the stellar rotation period, while in 7 cases it is about half that value, possibly due to the presence of X-ray-emitting structures at opposite stellar longitudes \cite{Flaccomio+2005COUP}. 
The observed modulation indicates that the X-ray-emitting regions are distributed non homogeneously in longitude and, usually, do not extend to distances significantly larger than the stellar radius. Modulation is observed in stars with saturated activity levels ($L_X/L_{Bol} \rm \sim 10^{-3}$) showing that saturation is not due to the filling of the stellar surface with X-ray-emitting regions. 

COUP has reached a sensitivity better than $L_X \rm = 10^{27}$ erg/sec providing a large unbiased sample of YSOs down to BDs and it is more than 95\% complete \cite{Preibisch+2005ApJSCOUP_Origin}. It arguably still is the best sample to study age-activity relationship in a very reliable way allowing, for the first time, to consider mass-stratified subsamples. The evolution of low-mass stars/YSOs magnetic activity -- of which $L_X$ is a proxy -- provides, together with the age-rotation relation, the observational constraints on stellar magnetic dynamo models and theories. For the range of ONC member ages  ~0.1-10 Myr \cite{PallaStahler1999ApJ}, a mild decay of $L_X$ with stellar age $\tau$ roughly as $L_X \sim \tau^{-1/3}$ has been found \cite{PreibischFeigelson2005ApJS}.
Comparing ONC YSOs with main-sequence stars up to ${\rm Log}~\tau [yr] = 9.5$, in the $\rm 0.5 { M_{\astrosun} < M < 1.2 M_{\astrosun}}$ mass range $L_X$ decays more rapidly as $L_X \sim \tau^{-0.75}$. Taking $L_X/L_{Bol}$ and the X-ray surface flux  $F_X$ as activity indicators, the decay is similar for the first 1-100 Myr but larger for older stars.  For masses in the 0.1 $\rm {M_{\astrosun}} < M < \rm 0.4 M_{\astrosun}$ range, i.e. M-type on MS, activity indicators have a different evolution. In the first 1-100 Myr a mild decrease of $L_X$, $L_X/L_{Bol}$ and $F_X$ has been found, followed by the three activity indicator decay over long timescales on the main sequence. 

Given also the confirmed lack of a rotation-activity relationship, the activity-age decay characterizing the evolution of solar-mass stars cannot be attributed to rotational deceleration during the early epochs. For $\rm {0.5 M_{\astrosun}} <M< \rm 1.2 M_{\astrosun}$ the results may be explained with a combination of tachocline and distributed convective dynamos at work up to $\sim$ 10 Myr, while for $\rm {0.1 M_{\astrosun}} <M< \rm 0.4 M_{\astrosun}$ the results are consistent with the convective dynamo dominance during the entire evolutive history.

\subsubsection{CSI-2264: Unveiling circumstellar disks with simultaneous multi-wavelength variability studies}
\label{sec:5.2.5}
The "Coordinated Synoptic Investigation of NGC 2264" (CSI-2264), lead by G. Micela and J. Stauffer, during which in December 2011, {\em simultaneous} observations of a large sample of YSOs in NGC~2264 have been obtained with three space-borne telescopes, \textit{Chandra} (X-rays), \textit{CoRoT} (optical), and \textit{Spitzer} (mIR) (e.g. \cite{Cody+2013,Cody+2014,Stauffer+2016}). CSI-2264 is the only program of this kind that has so far been done\footnote{Few selected interesting individual sources (e.g. AB Dor, V4046 Sgr, V2129 Oph, etc.), some of which are YSOs, have also been the subject of simultaneous multi-wavelength observations.}.  

The main mechanisms responsible for the YSO X-ray variability besides flares are variable extinction (e.g. warped rotating disk), unsteady accretion, and rotational modulation due to both hot, accretion induced, and cold photospheric spots and X-ray-active regions. In disk bearing YSOs, this variability is related to the morphology of the inner circumstellar region 
($ \leq 0.1$ AU)
and that of the photosphere and corona, all impossible to be spatially resolved with present-day techniques.  Thanks to the CSI-2264 data the X-ray spectral properties during optical bursts and dips to unveil
the nature of these phenomena have been studied \cite{Guarcello+2017}. 
The simultaneous CoRoT and \textit{Chandra/ACIS-I} observations have been analysed to search for coherent optical 
and X-ray flux variability. In YSOs with evidence of variable extinction, looking for a simultaneous increase of optical extinction and X-ray absorption during 
the optical dips; in YSOs showing accretion bursts, searching for soft X-ray emission and increasing X-ray absorption during the bursts. In 38\% of the 24 YSOs with optical dips, a simultaneous 
increase of X-ray absorption and optical extinction have been found. In seven dips, it was possible to calculate the $N_{H}/A_{V}$ ratio in order to infer the composition of the obscuring material. In 25\% of the 
20 YSOs with optical accretion bursts, increasing soft X-ray emission during the bursts arguably associated to the emission of accreting gas has been reported. Since favourable geometric configurations are required, these phenomena have been observed only in a fraction of YSOs with dips and bursts. The proposed scenario is that the observed variable absorption during the dips is mainly due to dust-free material in accretion streams. In YSOs with accretion bursts a larger soft X-ray spectral component has been, on average, observed while it has not been seen in weakly accreting YSOs.

\subsection{X-rays from Class~0 YSOs}
\label{sec:5.3}
Notwithstanding large observational efforts (see Sect. \ref{sec:4}) for long time there has been no clear evidence of X-ray emission from the
Class~0 YSO.
While X-rays are quite
penetrating  Class~0 sources can be subject to extinction up to
100 magnitudes and even higher preventing the escape of any X-rays. The most constraining upper bound to Class~0 $L_X$ was 
$\rm \sim 4 \times 10^{29}$ erg/s \cite{Giardino+2007aAA} that, however, is about two dex higher than the X-ray luminosity of active
Sun.  After several controversial reports more firm evidences have been recently reached.
By assembling the deepest and most, at the time, complete photometric catalogue of objects in the ONC region from the UV to 8 $\rm \mu$m a sample of high-probability YSOs members from Class~0 to Class~III have been selected and the properties of their X-ray emission has been studied concluding that Class 0-Ia YSOs are significantly less X-ray luminous than the more evolved Class II YSOs with $M$ > 0.5 $\rm M_{\astrosun}$ \cite{Prisinzano+2008ApJ}. Data are consistent with the onset of X-ray emission occurring at a very early stage. Class 0-I have X-ray spectral properties similar to those of the Class II and III, except for a larger absorption likely due to gas in the envelope or disk of the protostars.

Recently it has been reported that HOPS 383, a bona-fide Class~0 YSO, has been detected above 2 keV with \textit{Chandra} 
during a flare of a duration of $\sim$ 3.3 hr \cite{Grosso+2020AandA}. In the 2--8 keV bandpass the YSO peak $L_X$ has reached $\rm \sim 4 \times 10^{31}$ erg/s, at least 10 times higher that its undetected "quiescent" emission. 
The 28 collected counts have secured the detection, but required a careful analysis to extract spectral information. 
The flare spectrum is highly absorbed ($N_H \rm \sim  7 \times 10^{23} cm^{-2}$), the quality of the spectral fit does slightly improve when a Fe 6.4 keV emission line with an equivalent width of $\sim$ 1.1 keV is added, suggesting emission component arising from neutral or low-ionization iron. This results, if representative of all Class~0 YSOs, points toward an early development of 
the intense X-ray emission of YSOs and, as a result, the X-ray emission likely starts to regulate very early 
the further evolution of star formation process, for example by
determining the effectiveness of ambipolar diffusion (see Sect. \ref{sec:5.6}). Since the Class~0 data are still too sparse further deep observations with current or future X-ray observatories are needed.

\subsection{The YSO Flares: nature and effects on circumstellar disks}
\label{sec:5.4}

The issue of the existence of star-disk interconnecting flaring arches has been matter of debate over the last two decades. The issue is particularly relevant since, 
depending on the actual occurrence rate of those large flares, they can affect the early evolution of circumstellar disks with far reaching effects even on the formation of planetary systems. Detailed MHD model investigation has also shown that the flare location (and geometry) likely plays a key role (e.g. \cite{Colombo+2019AA}) in disk evolution.

\begin{figure}[t]
\centering
\vspace{-0.2cm}
\includegraphics[width=5.5cm]{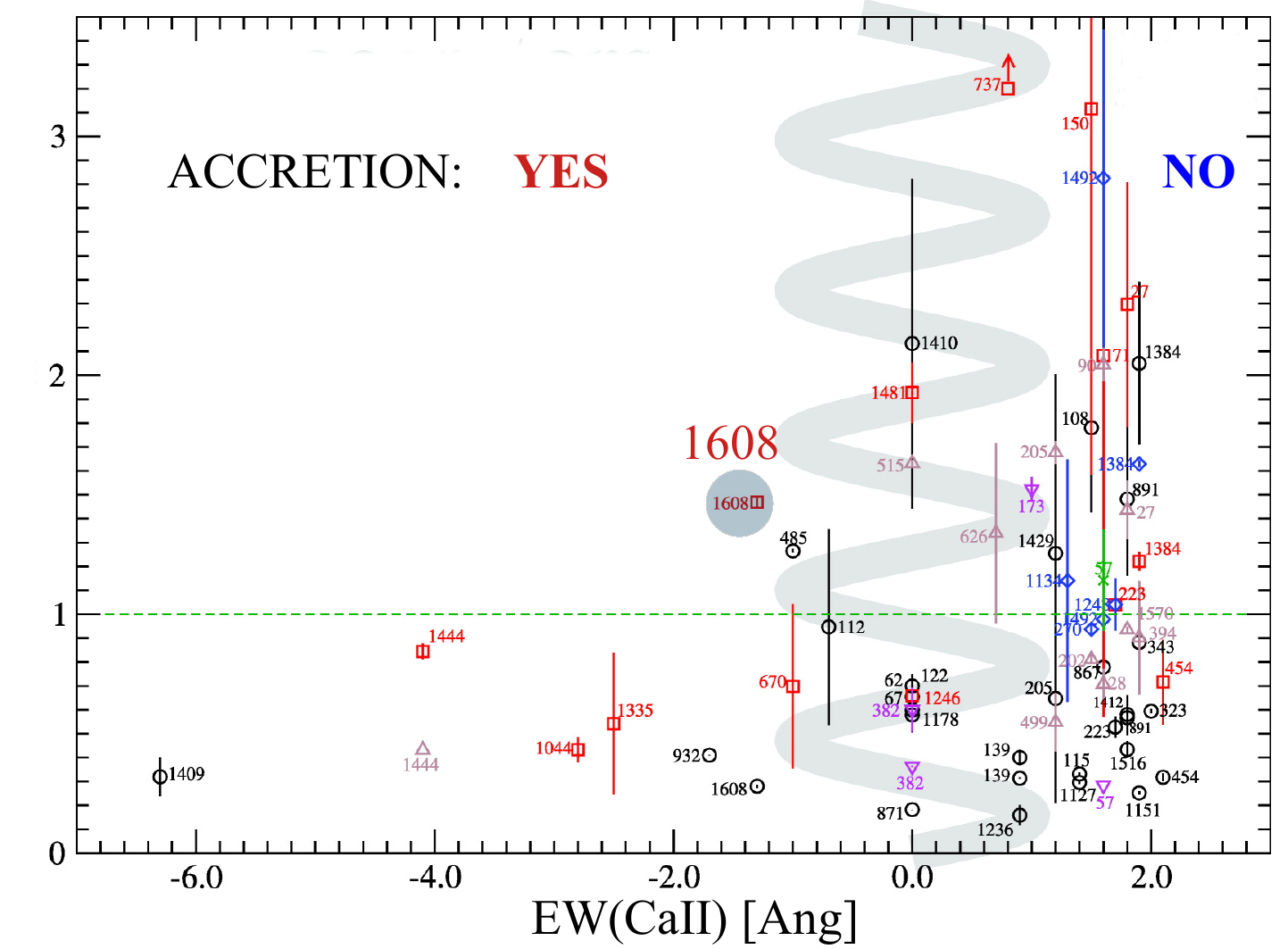}
\includegraphics[width=5.5cm]{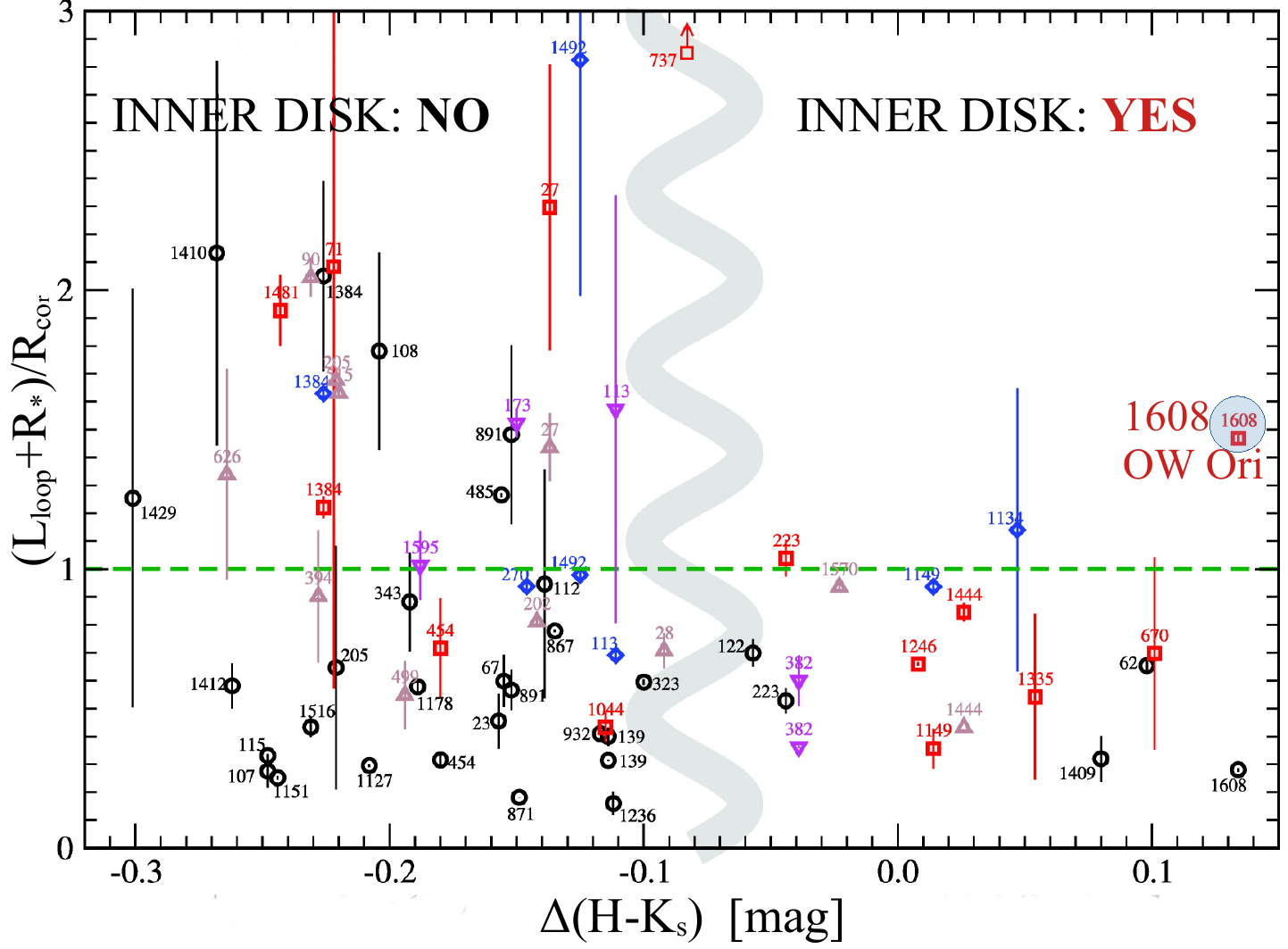}

\caption{(left) Scatter plot of $(L_{loop}+R_{*})/R_{cor}$ vs. excess of ($H-K_{S}$) color, an indicator of the presence of circumstellar disk. 
(right) Scatter plot of $(L_{loop}+R_{*})/R_{cor}$ vs. the EW of CaII triplet, an indicator of accretion process. 
In both panels the vertical wavy gray curve marks the transition to accreting/inner disks.
COUP 1608, that is outlined, clearly behaves very differently from the rest of (most of) 
the other YSOs. COUP 43 and COUP 332, for which all the different analysis techniques discussed in the text have derived long flaring structure, are not shown in the plots for lack of period and/or EW data
(adapted from \cite{Getman+2008b}).
}
\label{fig:4}
\end{figure}
 
A flare is a manifestation of the impulsive release of energy in a tenuous plasma confined in a "magnetic bottle" that loses energy by optically thin radiation and by efficient thermal conduction to the chromosphere (cf. the extensive discussion in \cite{Reale2007,Reale2014}). 
The magnetic confinement determines the typical flare light curve shape\footnote{The shape would be very different in the case of unconfined plasma} characterized by a very rapid increase of the emission (due to the rapid heating of the plasma) 
followed by a slow, almost exponential decay (due to the cooling by thermal conduction and radiative losses).
Most of the observed flares conform to very general conditions \cite{Reale2007} such that the time evolution of the emission and of the peak temperature, as well as the cooling, traced in the ${\rm Log}~(T) - \rm{Log}~(\sqrt{EM})$ diagram provide a ``direct'' estimate of the length of the flaring magnetic structure. Adopting this robust interpretative framework, 
the analysis of a sample of 32 large COUP flares has reached the conclusion that about 10 of them show flare decay times of about 1-2 days (cf. the published light-curves) and the flaring structure has a length of 3 -- 20 stellar radii, R$_*$,\cite{Favata+2005COUP}. 
Such a length structures have never been found in the many analogous analyses of flares in more evolved normal stars.
Structures of such extent, if anchored on the stellar surface, should suffer of likely stability problem due to the 
centrifugal force  since 1-2 Myr old YSOs are fast rotators with rotation period, P $\sim$ 3-6 days. The study of
the coronal loop equilibrium condition in fast rotating stars  assuming a magnetic field with potential configuration has shown that loops longer than $\sim$ 3 times the stellar radius are a possible stable solution in {\em disk-less} YSOs only assuming that the wind cooperates to support them \cite{Aarno+2012MNRAS}. In other configurations long loops anchored only on the star would be very likely ripped open during a long flare whose duration is of the order of rotational period (for further details cf. the discussion section of \cite{Favata+2005COUP}).
As a solution to this problem, since the co-rotation radius of those YSOs is typically at 4-5 R$_*$, it has been suggested that the loop, on which 
the flare occurs, is connecting the star and the disk (at the co-rotation radius). The existence of such magnetic ``funnels'' in class I-II YSOs is postulated by magneto-spheric accretion scenario (e.g. \cite{Hartmann2008book} and references therein).

In DROXO, in seven YSOs intense flares have been observed \cite{Flaccomio+2009} and the length of 
the flaring loops has been derived, in 2 cases (DROXO~63 and DROXO~67), the flaring loops are several stellar radii long. 
The fraction of the very long flaring structures in $\rho$ Oph is similar to the one observed in Orion, namely $\sim$ 30\% .    

By adopting a novel spectral analysis technique that avoids non-linear parametric modelling, 
the full set of 216 COUP flares on 161 YSOs have been analysed \cite{Getman+2008a,Getman+2008b} deriving the length of the flaring loop, $L_{loop}$. Only 98 of them, classified with the available IR photometry, have been retained in the further analysis.
Based on estimation of the stellar radius, $R_{*}$, and disk keplerian co-rotation radius, $R_{cor}$\footnote{The distance from stellar surface at which the angular velocity of disk equates that at stellar surface} (when the rotational period is known), 
the scatter plots of
$(L_{loop}+R_{*})/R_{cor}$ as a function of indicators (when available) of the presence of circumstellar disk or 
of on-going accretion process has been build for a final subsample of $\sim$ 70 flares (see Fig. \ref{fig:4})
concluding that: 1) circumstellar disks have no effect on flare morphology; 2) circumstellar disks may truncate PMS magneto-spheres, i.e., $(L_{loop}+R_{*})/R_{cor} < 1 $.
Moreover the study has found that circumstellar disks are unrelated to flare energetics and super-hot (> 100 MK) "non-standard" flares do occur in accreting YSOs (in agreement with \cite{Favata+2005COUP}).
Points 1) and 2) are at odd with the likely existence of star-disk interconnecting magnetically confined flaring structures \cite{Favata+2005COUP}.
 
A detailed analysis of the 68 flares on 65 YSOs {\em simultaneously} observed in the CSI-2264 program comparing the light curves obtained in X-rays (with \textit{Chandra}), 
in optical (with COROT), and in the infrared (with Spitzer) has been performed \cite{Flaccomio+2018AA}. The simultaneity of those observations -- crucial for studying fast variability phenomena like flares -- makes this really a {\em unique} dataset that includes the first, and still unique, simultaneous X-ray and IR observations of flare in a Class~I YSO. Over the studied sample
the flare released energy in the X-ray bandpass ranges from $\rm \sim 6 \times 10^{33}$ to $\rm \sim 2 \times 10^{36}$ erg. Some firm conclusions have been drawn, namely, 
i) the flare peak luminosity measured in the optical, IR and X-ray band-passes are tightly correlated, with a small (.3 dex) scatter, a similar relation (with a similar amplitude scatter) 
holds also for the flare energy released in the optical, IR and 
X-ray band-passes, the relationships hold over 3 orders of magnitude; 
ii) the flare energy emitted in "soft" X-rays is about 10\% to 20\% of the flare energy emitted in the optical band; 
iii) the flare energies are up to $\sim 5$ decades higher than those of
the brightest solar flares, and the simple extrapolation of solar flares to this extreme regime requires some cautions. As an example, the data indicate that the flare photospheric temperature 
is significantly lower than $10^4$ K, that is the typical solar value; 
iv) the occurrence of X-ray flares without optical counterparts can be due to flaring loop geometry since optical/IR emission comes from loop footprints whereas the X-rays are likely emitted along most of the loop. Based on a simple model the fraction of flares without optical counterparts ranges from 19\% to 48\%. 
v) there is evidence of a strong IR excesses for flares in stars with circumstellar disks: likely a result of the direct response (heating) of the inner disk to the flare. This is still the more {\em direct} observational signature of the interaction between flare emission and disks.

Taking advantage of the MYStIX and SFiNCs data products a large sample of powerful YSO flares has been recognized and studied \cite{Getman+2021aApJ, Getman+2021arXiv210608262G} deriving the peak X-ray luminosity ($L_{X,p}$) and released energy in the X-ray bandpass ($E_X$). The sample has been subdivided in 636 super-flares (34 <Log $E_X$ [erg] < 36.2) and 450 mega-flares (Log $E_X$ [erg] > 36.2). This latter subsample is the most complete of the two, the "best-fitting" flare energy distribution is $ \rm{d} N/\rm{d} E_X \propto {E_X}^{-1.95 \pm 0.07}$ over the entire mass range. The derived slope is in agreement with previous analyses \cite{Stelzer+2007AA, Caramazza+2007AA, Albacete+2007}. 
The occurrence rate is 1.7 [+1,-0.6] flares/YSO/year over the entire mass range, reduces to 0.3 [+0.2,-0.1] for $M$ < 1 $\rm M_{\astrosun}$ and is 11.0 [+6.4,-4.1] for $M$ > 1 $\rm M_{\astrosun}$.
From the derived flare intensity distribution and mass-dependent occurrence rate the contribution of observed mega-flares to the YSO X-ray fluence has been estimated to typically be 8-19\% and to decrease with YSO mass. This contribution could be even more conspicuous if much bigger flares sustained by extremely strong magnetic field would exist. However, so far, the maximum measured averaged YSO magnetic field reaches $\sim$ 3.3 kG (cf. \cite{Sokal+2020ApJ} and references therein); in a region of 0.1 $R_*$ the associated stored magnetic energy would be $\rm \sim 10^{39}$ erg, not far from the maximum measured energy of mega-flares.

\begin{figure}[t]
\vspace{-0.2cm}
\centering
\includegraphics[width=8cm]{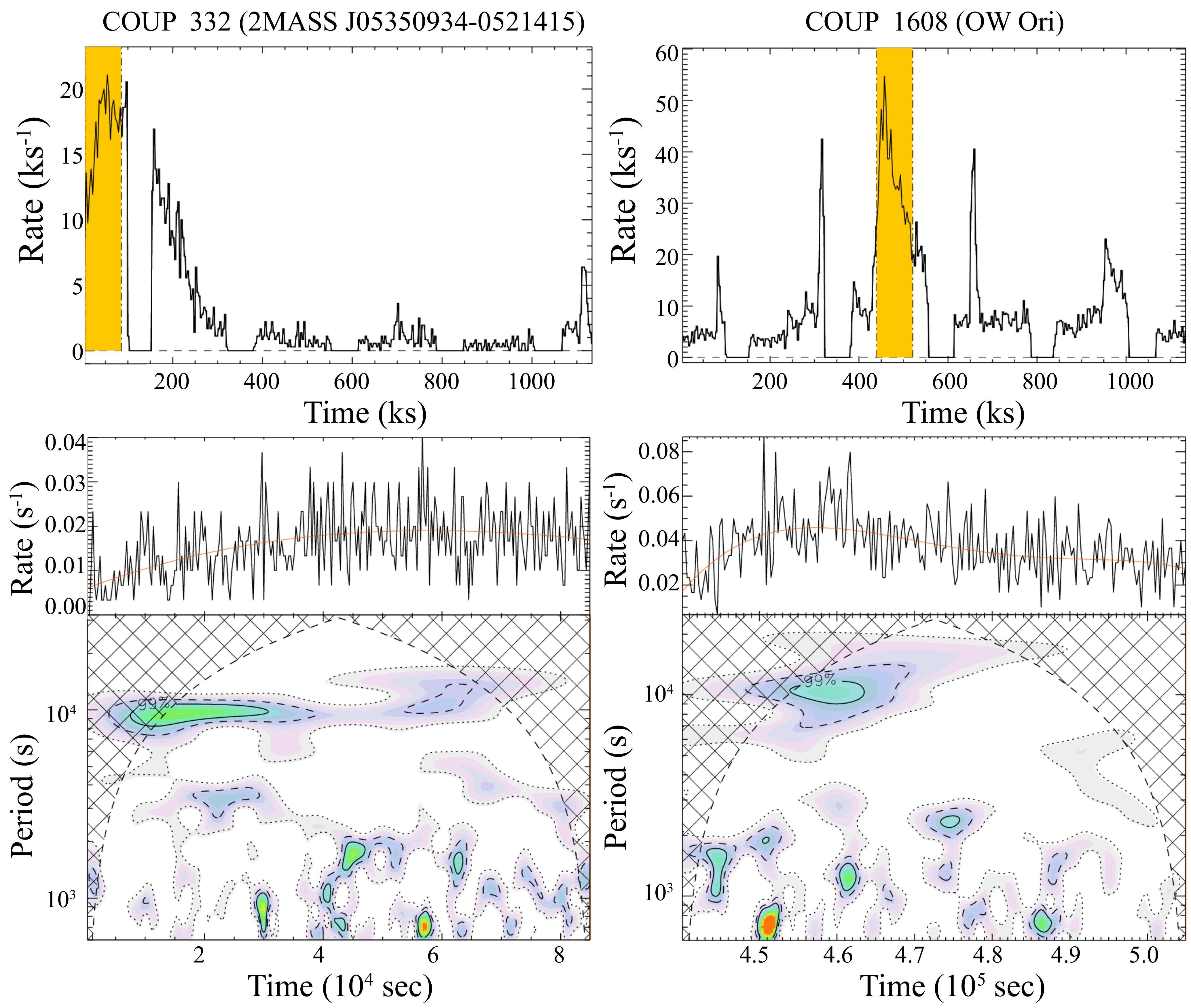}
\caption{The data and the summary of the wavelet analysis results for two COUP big flares. Time is measured from the beginning of the observation; the analysed flare data segment is highlighted in yellow. The central panel shows
the analysed data segment after subtraction of the running average, while the bottom panel shows the intensity curve as function of period and time. The 
statistical significance maps are also shown, they allow to discriminate the statistically 
significant period and the duration of the periodic signal. 
The dashed outer region marks the so-called ``cone of influence'', inside which the analysis provides meaningful results. 
The right and left panels are for COUP 1608 (OW Ori) and for COUP 332 (2MASS J05350934-0521415), respectively (adapted from \cite{Lopez-Santiago+2016}).}
\label{fig:5}
\end{figure}

By adopting the near and mid-IR photometric excess\footnote{More specifically the infrared spectral energy distribution slope $\alpha_{IRAC}$ with the value of -1.9 separating disk-less and disk-bearing YSOs \cite{Richert+2018MNRAS}.} as a proxy of the presence of the (gaseous) inner circumstellar disk the distributions of $L_{X,p}$ or $E_X$ of disk-less and disk-bearing YSOs have shown to be statistically indistinguishable confirming previous results on smaller samples \cite{Stelzer+2007AA, Flaccomio+2012, Flaccomio+2018AA}. Since big flares have been reported both in disk-less and disk-bearing YSOs this is not too surprising since duration, 
$L_{X,p}$ and $E_X$ are determined by the same physics governing all flares, irrespective of the YSO nature. It is worth to remember that the subdivision in disk-less and disk-bearing is likely a simplification as the ample range of accretion rates measured among  Orion YSOs clearly shows \cite{Manara+2012ApJ}.
Possibly more relevant it would be a comparative study of the (big) flare occurrence rates between the accreting and non-accreting YSO samples. In the Taurus-Auriga region, XEST data have shown a marginal hint that large amplitude and fast rise flares are more frequent on cTTS than on wTTS\footnote{The classification scheme adopted in XEST is predominantly based on accretion signature \cite{Guedel+2007XEST}.} (31 $\pm$ 7\% vs. 22 $\pm$ 7\%) \cite{Stelzer+2007AA}. A similar effect has also been found with a time resolved analysis of all available COUP data (cf. Fig. 6 of \cite{Flaccomio+2012}), but the differences in flare detectability among the fainter Class II and Class III YSO populations could bias the flare occurrence rates (cf.
Appendix A of \cite{Flaccomio+2012}) making difficult to draw firm conclusion. This same analysis has shown that 
disk-bearing stars are definitively more X-ray variable than disk-less ones, this could be explained as the effect of time-variable absorption by warped and rotating circumstellar disks, but some other effects due to disk presence cannot firmly be ruled out (at least in some of the YSOs).

The MYStIX/SFinCs flare study does not include the Carina complex and COUP nor CSI-2264. Let me stress that the COUP sample because of the time coverage, duration of continuous observations, accumulated count statistics and sample size and the CSI-2264 sample for its multi-wavelength simultaneous coverage remain the best ones to investigate the origin of (big) flares.  

Even in a systematic analysis of COUP data \cite{Getman+2008b} there are a few YSOs whose data can hardly be reconciled with the lack of any evidence of a possible effect of circumstellar disk, notably COUP~1608 and COUP~332 \cite{Lopez-Santiago+2016} and COUP~43 \cite{Reale+2018}, three of the about 10 COUP YSOs with long flaring loop \cite{Favata+2005COUP}.

\begin{figure}[t]
\includegraphics[width=7.5cm]{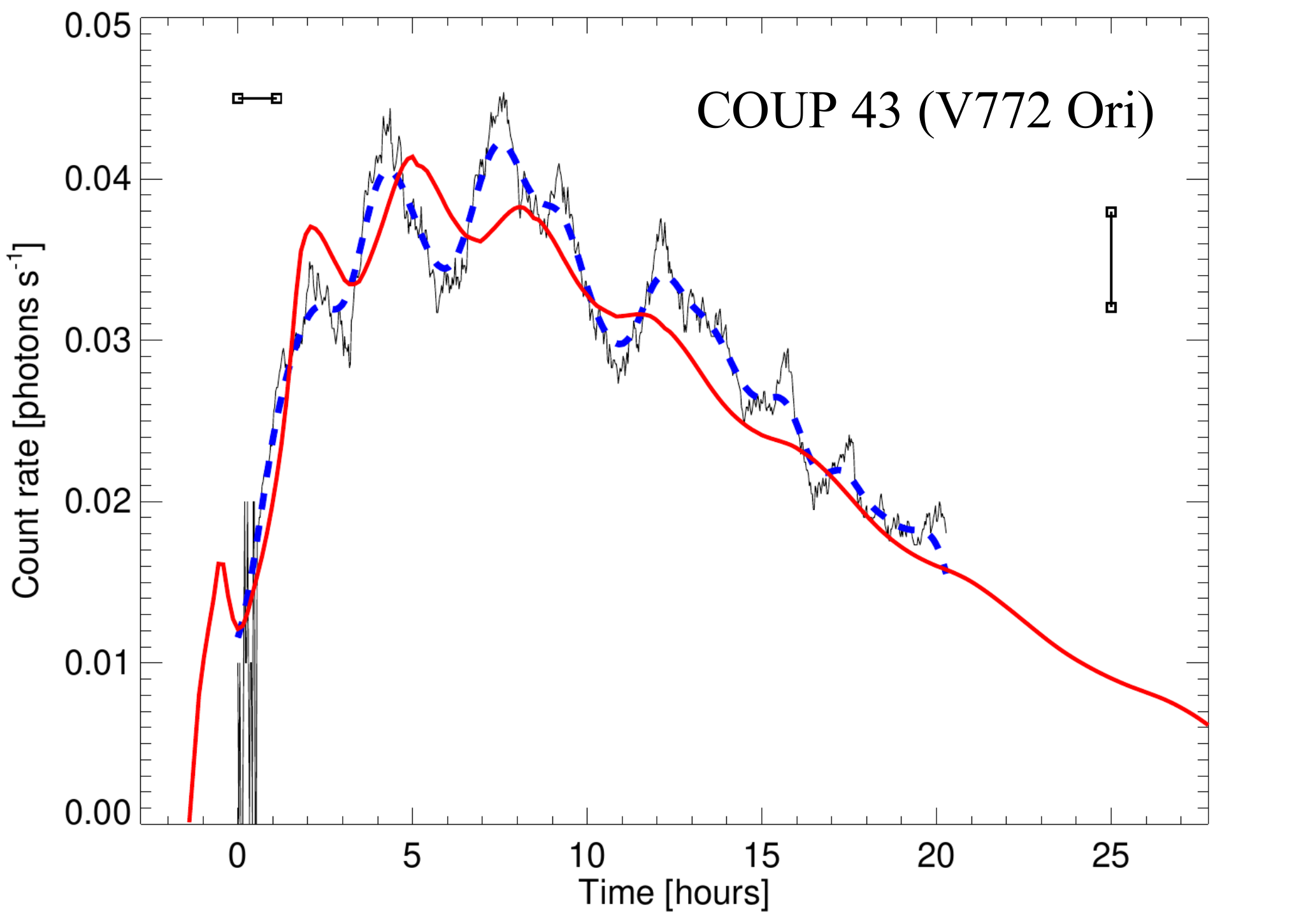}
\texttt{\sidecaption
\caption{\label{fig:6} The smoothed and running average subtracted light-curve of COUP 43 flare (black line) is compared with the hydrodynamical model 
synthesized light-curve in the case of a short duration heating pulse (blue dashed line) and of a long duration heating (red line) on the same long length flaring structure. 
Only a long loop with a length of about 10-20 $R_{\astrosun}$  and a short heating pulse can reproduce the observed oscillations both in intensity and period (adapted from \cite{Sciortino+2019AN}).}
}
\end{figure}

A novel analysis of the light curves based on the 
so-called Morlet wavelet (e.g. \cite{LopezSantiago2018} and references therein) has been applied to some 
of the aforementioned big COUP flares. 
This analysis has shown to be very powerful in detecting quasi-periodic oscillations occurring in some
stellar coronal flares.  On the basis of simple physical argument, it is possible to derive from measured period the length of the flaring structure where the oscillating X-ray emission comes from \cite{Lopez-Santiago+2016}.
In the case of some of the COUP big flares it has been shown the existence of oscillations during the flare decay phase allowing to determine the oscillation period \cite{Reale+2018}. More specifically the analysis has found
large-amplitude ($\sim$ 20\%), long-period ($\sim$ 3 hr) pulsations in the light curve of two day-long COUP flares. Detailed hydrodynamical modelling, including all the relevant physical effects, of the flares observed on COUP 43 (V772 Ori)
(shown in Fig. \ref{fig:6})
and COUP 1068 (OW Ori) shows that these pulsations track the sloshing of plasma along an elongated 
magnetic tube, triggered by a heat pulse whose duration ($\sim$ 1 hr) is much shorter that the sound crossing time along the loop. From this simple and robust modelling the authors have concluded that
the involved magnetic tubes are $\sim$ 20 solar radii long, and, very likely, connect the stars (near a polar region) with their surrounding disks (near the corotation radius). Such a geometry is consistent with the observed day-long stability of the flaring loop.

All those evidences taken together indicate that we still lack a clear understanding of the nature of those big (flaring) loops; even their stable existence in disk-less YSOs requires a specific modelling \cite{Aarno+2012MNRAS} involving the YSO winds; the extension to the case of disk-bearing YSOs should be properly investigated\footnote{A recent study of protoplanetary disk winds shows that both fast and slow winds are consistent with expectations from a thermal-magnetized disk wind model and are generally inconsistent with a purely thermal wind \cite{Ziyan+2021arXiv210711188X}}.
On the possible effect of circumstellar disks further studies with higher quality data on truly unbiased samples are needed; there is a large quantity of data, but their relevance/quality and/or absence of, often subtle, biases requires a critical analysis; a few points worth to consider are: i) the large flaring loops are a coronal phenomenon involving hot {\em gas}, the presence of dust is not really required, however the IR excess to distinguish disk-less and disk-bearing YSOs is sensitive to the presence of {\em dust}; ii) for the star-disk interaction it is likely relevant the accreting status of the disk (and not only its mere existence), something that should be evaluated by specific, and as "direct" as possible, accretion signatures; iii) subtle biases induced by the "selective" lack of accretion signature and/or rotational period should be carefully analysed.
The fact that most of the scaling laws of observed flares are similar and/or extensions of the solar case \cite{Getman+2021arXiv210608262G} is interesting, but not too much surprising since the physical nature of flares is always that of an explosion in a (magnetic) bottle with conductive and radiative losses. Perhaps more revealing could be the comparison of flare occurrence rates of well defined samples of (strongly) accreting and non/weakly-accreting YSOs.

\subsection{Circumstellar Disk Evolution and High-energy radiation}
\label{sec:5.5}
The issue of the effect of high-energy radiation on circumstellar disk evolution has been studied and debated. Disks around low-mass stars can be strongly perturbed or even destroyed before they form planets by: photoevaporation induced by incident X-ray and UV radiation (e.g. \cite{StorzerHollenbach1999ApJ,Picogna+2019MNRAS,Ercolano+2021MNRAS}) or gravitational interaction during close encounters with other cluster members (e.g. \cite{Pfalzner+2005ApJ, Vincke+2015AA}). 
Evaporating disks have been observed in the Orion Trapezium (e.g. \cite{Bally+2000AJ}), 
Cygnus OB2 \cite{Wright+2012ApJ, Guarcello+2014ApJ}), NGC~2244 \cite{Balog+2006ApJ}, 
NGC~1977 \cite{Kim+2016ApJ}, and Carina \cite{Mesa-Delgado+2016ApJ}. Indirect evidence supporting a 
fast erosion of proto-planetary disks in proximity of massive stars was deduced by the decline of the disk fraction observed close to massive stars or in regions with high local UV fields in massive clusters/associations such as: 
NGC~2244 \cite{Balog+2007ApJ}, using Spitzer data alone, and NGC~6611 \cite{Guarcello+2007AA, Guarcello+2009AA, Guarcello+2010AA} and Pismis~24 \cite{Fang+2012AA} using a multi-wavelength approach with a pivotal role played by \textit{Chandra} data. 
In contrast with these results, in the MYStIX massive cluster sample
no evidence supporting a lower disk fraction near massive stars was instead reported \cite{Richert+2015ApJ}, suggesting that evidence supporting the external disks photo-evaporation found by earlier studies was affected by selection effects. 
Subsequent \textit{Chandra}-enabled studies of e.g. NGC~6231 \cite{Damiani+2016AA}, 
Cygnus OB2 \cite{Guarcello+2016arXiv160501773G}, and Trumpler~14 and 16 \cite{ReiterParker2019MNRAS} have refuted this hypothesis 
supporting the evidence that the star-forming environment plays an important role in the survival and enrichment of 
proto-planetary discs and confirming disk photo-evaporation role. A recent study of Dolidze~25 suggests that disk evolution may be impacted by the environment; given the small number of O stars and the low stellar density, disk dispersal time scale is likely determined by cluster low metallicity rather than photo-evaporation or dynamical encounters \cite{Guarcello+2021AA}.

A signature of X-ray driven disk photo-evaporation due to the low-mass YSO emission is provided by the dependence of stellar accretion rates on X-ray luminosities, $L_X$; models predict that, in a coeval, similar mass sample, stars with higher $L_X$ should show, on average, lower accretion rates. By using the COUP derived $L_X$ \cite{Getman+2005COUP} and accretion derived from the Orion HST Treasury Program \cite{Manara+2012ApJ} it has been found evidence for a weak anti-correlation, as predicted by the models, even after correcting for the $L_X$ and accretion rate dependence on mass that could produce a spurious anti-correlation \cite{Flaischlen+2021AA}.

Since massive stars are the primary sources of X-ray/UV radiation and the chances of close encounters are high in dense stellar environments, both processes 
can be relevant depending on massive stars numbers and space density. The most extreme environment being a massive starburst region like Westerlund~1 (WD1). Thus, likely better evidence would be obtained testing proto-planetary
disk survival in such an extreme environment; this is really a challenging experiment, at the limit of the capability both of \textit{Chandra} and \textit{JWST}. It is one of the main objectives of the currently ongoing "Extented Westerlund One Chandra, and JWST, Survey" (EWOCS) \cite{Guarcello+2021jwst}.

\begin{figure}[t]
 \includegraphics[width=6cm]{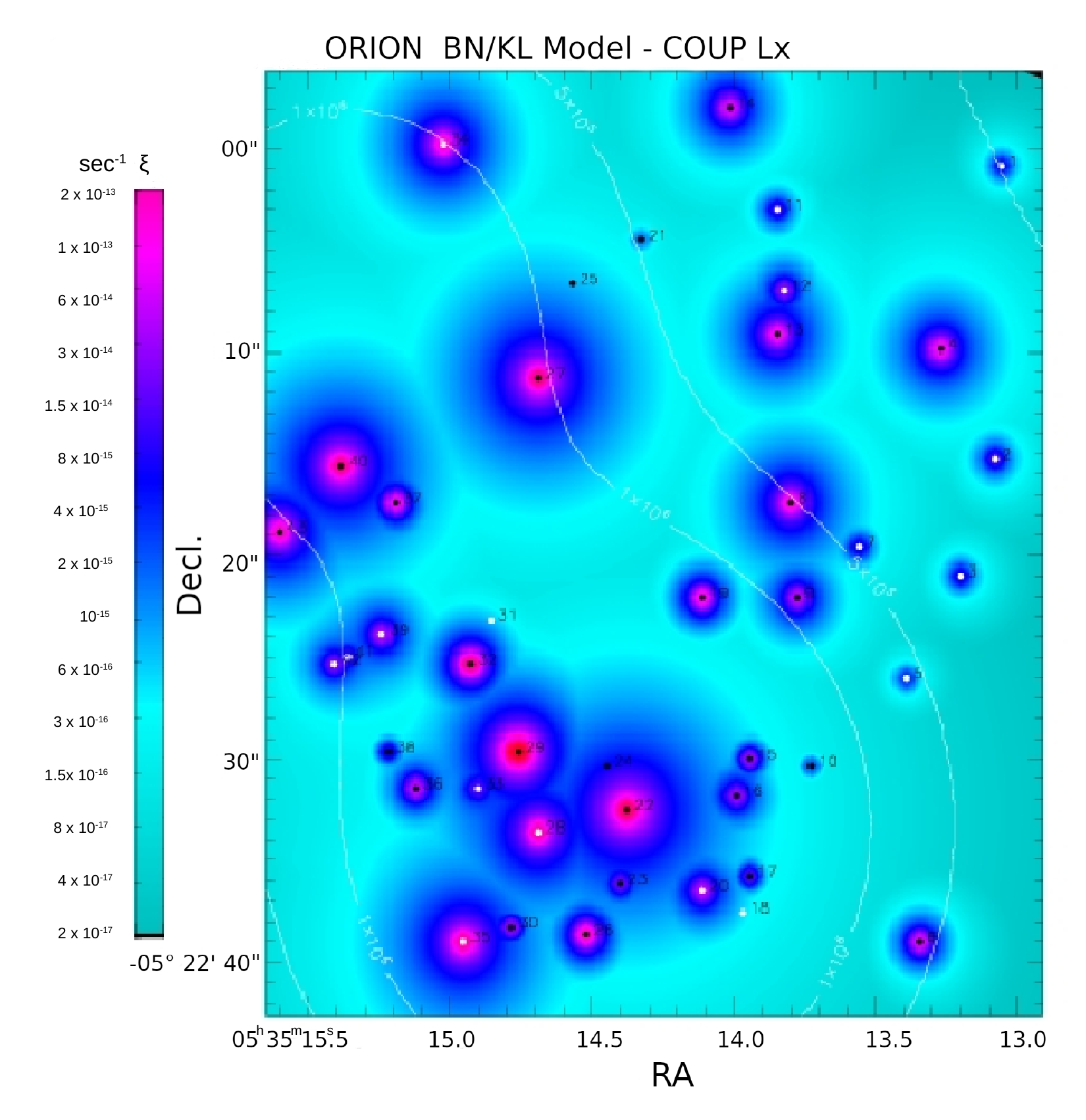}
 \texttt{\sidecaption
 \caption{A two dimensional projection of a model predicted
ionization rate for the BN-KL cloud core 
in the Orion Molecular Cloud~1 region based on COUP derived $L_X$.  Across the entire core the
ionization rate exceeds the value due to cosmic 
rays ($\rm 2 \times 10^{-17} sec^{-1}$) and around each of the 
embedded X-ray emitting YSOs develops a \textit{R\"ontgen} sphere where the 
X-ray induced
ionization rate is several orders of magnitude higher than the background level
(model data courtesy of the late F. Palla.)}
}
\label{fig:7}
\end{figure}

\subsection{YSO X-ray emission effects on small and large scales}
\label{sec:5.6}
The role of YSO X-rays as a major local ionization component has been modelled and discussed by \cite{LorenzaniPalla2001} who have introduced the concept of the \textit{R\"ontgen} sphere as the region around each YSO where the ionization rate due to the YSO X-rays is greater than the one due to cosmic rays. With the typical $L_X$ of YSOs the resulting ionization is 4-20 time higher than the value due to the cosmic rays. In SFRs like Orion and $\rho$ Oph the YSO space density is sufficient to affect regions of 1 pc size. Due to the greater ionization  the coupling of magnetic field to the cloud should increase and the cloud core collapse should take longer due to ambipolar diffusion. Hence the X-rays from YSOs could naturally slow down an accelerating star formation process.
More recently a much detailed treatment of the X-ray transfer and energy deposition into a gas with solar composition, with an accurate description of the electron cascade after the primary photoelectron energy deposition \cite{Locci+2018MNRAS} has confirmed that in molecular clouds embedding low-mass YSOs substantial volumes of gas are affected by ionization levels much higher than the cosmic-ray background ionization. 
In dense SFRs X-rays create an ionization network pervading densely the interstellar medium, and providing a natural feedback mechanism. These results support the idea that X-ray could be among the most relevant ingredients in the evolution of a variety of astronomical regions. Let me add that in the COUP data at least 3 YSOs have flare peak $L_X$ higher than the one found on HOPS~383 \cite{Wolk+2005COUP} and many more have been recently reported (see Sect. \ref{sec:5.4}) in many other SFRs. Big flares, with an estimated rate of $\sim$ 1.7 flares/YSO/year
\cite{Getman+2021aApJ}, are common among Class~I, and may be Class~0, YSOs. During the first 5 Myr of its evolution a solar mass YSO will generates about 10$^7$ big flares which will very likely play a role in the ionization of YSO circumstellar environments and disks. 

\section{A glance into the future}
\label{sec:6}
Making predictions is often rooted on biased opinions and is a slippery exercise, I do believe that astronomy is really driven by the advancement of observational capabilities; paradigm changing discoveries
have often been the outcome of the increase in discovery space fostered by new telescopes and instruments \cite{Fabbiano+2019arXiv190306634F}. 

Having this in mind there are still some obvious predictions: 1) the ongoing\footnote{Since March 2022 the survey is in hold due to international tensions; survey restarting date can hardly be predicted.} \textit{eRosita} all-sky survey ($\sim$ 30 times deeper than the RASS) will permit to investigate the YSO population in the nearby SFRs also thanks to the availability of \textit{GAIA} and IR data. A first example is provided by the study of the Sco-Cen OB association \cite{Schmitt+2022AA}. The foreseen eRosita pointed program will open the opportunity to perform long dedicated observations of interesting SFRs; if performed simultaneously at other wavelengths they will have a huge impact as we have learned with CSI-2264. Given the \textit{eRosita} angular resolution these studies should concentrate on rather nearby SFRs. Already at the distance of Orion, confusion, unless reaching the superb \textit{Chandra} angular resolution, could be a problem; 2) the \textit{XRISM} micro-calorimeter promises groundbreaking 7eV spectral resolution, namely a resolving power, $R \rm \sim 1000$ at Fe K$_{\alpha}$ 6.7 keV line(s) even if with limited angular resolution. The best targets for \textit{XRISM} YSO studies will be the nearby heavily obscured SFRs, e.g. $\rho$ Oph and R CrA. The strong Fe 6.7 keV and Fe fluorescent 6.4 keV lines, often found in YSO X-ray spectra, can diagnose plasma conditions and dynamics of the innermost region of Class~I and Class~II YSO where the free-fall and Keplerian velocities of matter and X-ray emitting plasma are expected to be a few hundreds km/s. At $\sim$ 7 keV \textit{Chandra}\footnote{The \textit{XMM-Newton} reflection grating bandpass extends only to about 2.4 keV.} high-resolution grating spectrometers lack the sensitivity and resolution to resolve such velocity shifts whose investigation at much lower energies ($\leqslant$ 1 keV) is hampered by absorbing circumstellar matter. For the first time \textit{XRISM} should be able to investigate a) the geometry of X-ray plasmas around the central object
from Doppler shift or broadening measurements of highly ionized Fe emission lines, b) the emitting plasma equilibrium state from multiple Fe emission lines, c) the dynamics of the innermost
accretion disk from Doppler shifts or broadening of the fluorescent Fe K$_{\alpha}$ line, d) the dynamics of plasma from Doppler shifts of emission lines during X-ray flares; 3) on a longer time-scale, with a possible launch around 2035, the under-study ESA L-class \textit{ATHENA} X-ray observatory with its large collecting area ($\sim$ 1.4 sq.m at 1 keV), spectral resolution (> 2.5 eV up to 7 keV), and long continuous observations will have a deep, transformational impact. As an example, it would be possible to study with a time resolution of a few ksec the possible time variations of the energy of the Fe K$_{\alpha}$ fluorescent line complex centroid of which we have found tantalizing evidence in a deep \textit{XMM-Newton/EPIC} observation \cite{Pillitteri+2019}. 

An analysis, admittedly somewhat outdated, of the \textit{ATHENA} capabilities on YSOs research is provided by \cite{Sciortino+2013arXiv1306.2333S}, two white papers discuss the foreseen synergy between \textit{ATHENA} and the ESO current and under construction telescopes \cite{Padovani+2017arXiv170506064P}, and those between \textit{ATHENA} and \textit{SKA} \cite{Cassano+2018arXiv180709080C}. They include, among the many, a discussion of multi-wavelength investigations of the star formation process, YSO physics, and evolution of circumstellar and proto-planetary disks.

In order to reach the cores of far away massive SFRs, as an (X-ray) astronomer I continue to envisage an X-ray observatory having a \textit{Chandra}-like resolution, a few square meter effective area, a $\sim$ 500 square arc-min field-of-view, a $\sim$ 1 eV spectral resolution, and a $\sim$ 1 millisecond time, when feasible -- substantial technological developments are required -- this will be a wonderful play field for the next generations of astronomers.

\begin{acknowledgement}
I sincerely thank the many colleagues whose long-standing dedication to astronomy have made possible the writing of this review, among the many I want to remember some of my younger colleagues in Palermo: Francesco Damiani, Ettore Flaccomio, Mario Giuseppe Guarcello, Ignazio Pillitteri and Loredana Prisinzano.  Some of the exposed ideas have benefited of the many pleasant friendly discussions I have had with the late Francesco Palla. This research has made use of NASA’s Astrophysics Data System.
\end{acknowledgement}

%

\def\ref@jnl#1{{#1 }}

\def\aj{\ref@jnl{AJ}}                   
\def\actaa{\ref@jnl{Acta Astron.}}      
\def\araa{\ref@jnl{ARA\&A}}             
\def\apj{\ref@jnl{ApJ}}                 
\def\apjl{\ref@jnl{ApJ}}                
\def\apjs{\ref@jnl{ApJS}}               
\def\ao{\ref@jnl{Appl.~Opt.}}           
\def\apss{\ref@jnl{Ap\&SS}}             
\def\aap{\ref@jnl{A\&A}}                
\def\aapr{\ref@jnl{A\&A~Rev.}}          
\def\aaps{\ref@jnl{A\&AS}}              
\def\azh{\ref@jnl{AZh}}                 
\def\baas{\ref@jnl{BAAS}}               
\def\bac{\ref@jnl{Bull. astr. Inst. Czechosl.}}
\def\caa{\ref@jnl{Chinese Astron. Astrophys.}}
\def\cjaa{\ref@jnl{Chinese J. Astron. Astrophys.}}
\def\icarus{\ref@jnl{Icarus}}           
\def\jcap{\ref@jnl{J. Cosmology Astropart. Phys.}}
\def\jrasc{\ref@jnl{JRASC}}             
\def\memras{\ref@jnl{MmRAS}}            
\def\mnras{\ref@jnl{MNRAS}}             
\def\na{\ref@jnl{New A}}                
\def\nar{\ref@jnl{New A Rev.}}          
\def\pra{\ref@jnl{Phys.~Rev.~A}}        
\def\prb{\ref@jnl{Phys.~Rev.~B}}        
\def\prc{\ref@jnl{Phys.~Rev.~C}}        
\def\prd{\ref@jnl{Phys.~Rev.~D}}        
\def\pre{\ref@jnl{Phys.~Rev.~E}}        
\def\prl{\ref@jnl{Phys.~Rev.~Lett.}}    
\def\pasa{\ref@jnl{PASA}}               
\def\pasp{\ref@jnl{PASP}}               
\def\pasj{\ref@jnl{PASJ}}               
\def\rmxaa{\ref@jnl{Rev. Mexicana Astron. Astrofis.}}%
\def\qjras{\ref@jnl{QJRAS}}             
\def\skytel{\ref@jnl{S\&T}}             
\def\solphys{\ref@jnl{Sol.~Phys.}}      
\def\sovast{\ref@jnl{Soviet~Ast.}}      
\def\ssr{\ref@jnl{Space~Sci.~Rev.}}     
\def\zap{\ref@jnl{ZAp}}                 
\def\nat{\ref@jnl{Nature}}              
\def\iaucirc{\ref@jnl{IAU~Circ.}}       
\def\aplett{\ref@jnl{Astrophys.~Lett.}} 
\def\apspr{\ref@jnl{Astrophys.~Space~Phys.~Res.}}
\def\bain{\ref@jnl{Bull.~Astron.~Inst.~Netherlands}} 
\def\fcp{\ref@jnl{Fund.~Cosmic~Phys.}}  
\def\gca{\ref@jnl{Geochim.~Cosmochim.~Acta}}   
\def\grl{\ref@jnl{Geophys.~Res.~Lett.}} 
\def\jcp{\ref@jnl{J.~Chem.~Phys.}}      
\def\jgr{\ref@jnl{J.~Geophys.~Res.}}    
\def\jqsrt{\ref@jnl{J.~Quant.~Spec.~Radiat.~Transf.}}
\def\memsai{\ref@jnl{Mem.~Soc.~Astron.~Italiana}}
\def\nphysa{\ref@jnl{Nucl.~Phys.~A}}   
\def\physrep{\ref@jnl{Phys.~Rep.}}   
\def\physscr{\ref@jnl{Phys.~Scr}}   
\def\planss{\ref@jnl{Planet.~Space~Sci.}}   
\def\procspie{\ref@jnl{Proc.~SPIE}}   

\let\astap=\aap
\let\apjlett=\apjl
\let\apjsupp=\apjs
\let\applopt=\ao

%
%

\end{document}